\newcommand{\diracslash}[1]{#1\llap{/\kern2pt}}

\newcommand{\be}{\begin{equation}}
\newcommand{\ee}{\end{equation}}
\newcommand{\bea}{\begin{eqnarray}}
\newcommand{\eea}{\end{eqnarray}}
\newcommand{\ba}[1]{\begin{array}{#1}}
\newcommand{\ea}{\end{array}}

\newcommand{\bt}{\begin{tabular}}
\newcommand{\et}{\end{tabular}}

\newcommand{\beas}{\begin{eqnarray*}}
\newcommand{\eeas}{\end{eqnarray*}}

\documentclass[preprint,prd,aps,floats,nofootinbib,floatfix]{revtex4}

\DeclareSymbolFont{rsfs}{U}{rsfs}{m}{n}
\DeclareSymbolFontAlphabet{\mathrsfs}{rsfs}
\usepackage{graphicx}
\usepackage{multirow}

\usepackage{graphicx}

\begin{document}

\title{Heavy Vector and Axial-Vector Mesons in Asymmetric Strange Hadronic  Matter}
\author{Arvind Kumar}
\email{iitd.arvind@gmail.com, kumara@nitj.ac.in}
\affiliation{Department of Physics, Dr. B R Ambedkar National Institute of Technology Jalandhar, 
 Jalandhar -- 144011,Punjab, India}
 \author{Rahul Chhabra}
\email{rahulchhabra@ymail.com,}
\affiliation{Department of Physics, Dr. B R Ambedkar National Institute of Technology Jalandhar, 
 Jalandhar -- 144011,Punjab, India}

\def\be{\begin{equation}}
\def\ee{\end{equation}}
\def\bearr{\begin{eqnarray}}
\def\eearr{\end{eqnarray}}
\def\zbf#1{{\bf {#1}}}
\def\bfm#1{\mbox{\boldmath $#1$}}
\def\hf{\frac{1}{2}}
\def\kp{\zbf k+\frac{\zbf q}{2}}
\def\km{-\zbf k+\frac{\zbf q}{2}}
\def\hwo{\hat\omega_1}
\def\hwt{\hat\omega_2}

\begin{abstract}

We  calculate the effects of finite density   of 
isospin asymmetric strange
hadronic matter, for different strangeness fractions,
on the in-medium properties of 
vector  $\left( D^{\ast}, D_{s}^{\ast}, B^{\ast}, B_{s}^{\ast}\right)$ 
and axial-vector  $\left( D_{1}, D_{1s}, B_{1}, B_{1s}\right)$ mesons
using chiral hadronic SU(3) model and QCD sum rules.
We focus on the evaluation of in-medium mass-shift and shift of decay constant
of above  vector and axial vector mesons.
In QCD sum rule approach the properties e.g. masses and decay constants
of vector and axial vector mesons
are written in terms of quark and gluon condensates.
These quarks and gluon condensates  are evaluated in the
present work using chiral SU(3) model
through the medium modification of scalar-isoscalar fields
$\sigma$ and $\zeta$, the scalar-isovector field
 $\delta$ and scalar dilaton field $\chi$ 
in  strange hadronic medium which 
includes both nucleons as well as hyperons.
As we shall see in detail the masses and decay constants
of heavy vector and axial vector mesons 
are affected significantly due to isospin asymmetry 
and  strangeness fraction of the medium
and these modifications may influence
the experimental observables
produced in heavy ion collision experiments. 
The results of present investigations of in-medium properties of
vector and axial-vector mesons at
finite density of strange hadronic medium may be helpful
for understanding the experimental data from heavy-ion collision experiments
in-particular for the 
Compressed Baryonic Matter (CBM) experiment of FAIR facility at
GSI, Germany. 

\textbf{Keywords:} Dense hadronic matter, strangeness fraction,
 heavy-ion collisions, effective chiral model, QCD sum rules,  heavy mesons.

PACS numbers : -14.40.Lb ,-14.40.Nd,13.75.Lb
\end{abstract}

\maketitle
\section{Introduction}
The aim of 
relativistic heavy-ion collision experiments
is to explore the different phases of QCD phase diagram
so as to understand the underlying strong
interaction physics of
Quantum Chromodynamics. The different regions of QCD phase
diagram can be explored by varying the beam energy in the
high energy heavy-ion collision experiments.
The nucleus-nucleus collisions at the Relativistic Heavy Ion Collider (RHIC)
and Large Hadron Collider (LHC)
experiments explore the region of the QCD phase
diagram at low baryonic densities and high temperatures.
However, the objective of the Compressed Baryonic Matter (CBM)
experiment of FAIR project at GSI, Germany is
 complementary to RHIC and LHC experiments.
It aims to study the region of phase diagram at high baryonic density and
moderate temperature. In nature these kind of phases may
exist in astrophysical compact objects e.g. in neutron
stars. Among the many different observable  which may be produced
in CBM experiment the one of them may be the production of mesons having
charm quark or antiquark.
Experimentally, charm meson spectroscopy as well as their
in-medium properties are also of interest from
the point of view of PANDA experiment of FAIR project
where $\bar{p}A$ collisions will be performed.
The possibility of production of open or hidden charm mesons
motivate the theoretical
physicist to study the properties of these mesons in
dense nuclear matter. The discovery
of many open or hidden charm
or bottom  mesons at CLEO, Belle or BABAR experiments \cite{cleo,bele,babar}
attract the attentions of theoretical groups to study the
properties of these mesons.
 
 Quark meson coupling model \cite{qmc1},
coupled channel approach \citep{tolos2,tolos3,tolos4}, 
QCD sum rules \cite{higler1,haya1,wang1,wang2} or chiral hadronic 
models \cite{paper3,amarind,amdmeson,amarvind,amavstranged} 
etc are among the many theoretical approaches
used for investigating the in-medium properties
of hadrons. 
The theoretical investigations of open or hidden charmed meson properties
at finite density or temperature of the nuclear matter
may help us in understanding their 
production rates, decay constants, decay widths etc in
heavy-ion collision experiments.
The study of open charm D mesons
also help in understanding the phenomenon of $J/\psi$ suppression
produced in heavy-ion collisions. The higher charmonium states are considered
as major source of $J/\psi$  mesons. However, if the D mesons 
undergo mass drop in the nuclear matter and the in-medium
mass of $D\bar{D}$ pairs falls below the threshold value of 
excited charmonium states then these states can also
decay to $D\bar{D}$ pairs and may cause a decrease in the yield of
$J/\psi$ mesons.
 In ref. \cite{amavstranged} the 
pseudoscalar $D$ and $\bar{D}$ mesons were investigated by generalizing the chiral
SU(3) model to SU(4) and the in-medium
masses were calculated at finite density of nuclear  and strange hadronic
matter.  The masses of $D$ and $\bar{D}$ mesons were found to be sensitive to
the density as well as the strangeness fraction of
the hadronic medium.
The chiral hadronic model is recently also generalized to SU(5) sector
so as to investigate the effects of density and  strangeness fractions
on the in-medium masses of open bottom mesons $B, \bar{B}$ and $B_{S}$.
In ref.\cite{haya1} the Borel transformed QCD sum rules 
were used to study the pseudoscalar D meson mass modifications
in the nuclear medium.  
The mass splitting of $D$  and $\bar{D}$ mesons was investigated
using QCD sum rules  in \citep{higler1}. The study of medium modifications of
$D$ and $\bar{D}$ mesons may help us in
understanding the possibility of formation of charmed mesic nuclei. The charmed mesic nuclei are
the bound states of charm mesons and nucleon formed through strong interactions.
In ref. \citep{qmc1} the mean field potentials of $D$ and $\bar{D}$ mesons
were calculated using the quark meson coupling (QMC) model 
under local density approximation and the possibilities of the
formation of  bound states of $D^{-}$, $D^{0}$  and $\bar{D^{0}}$
mesons with Pb(208) were examined. The properties of charmed mesons in the nuclei
had also been studied using the unitary meson-baryon coupled
channel approach and incorporating the heavy-quark spin symmetry \citep{tolosbnd,tolos5}.

The in-medium mass modifications of scalar, vector and axial vector heavy
$D$ and $B$ mesons were investigated using
 the QCD sum rules in the nuclear matter in \cite{wang1,wang2}.
 Recently we studied the mass-modifications of 
 scalar, vector and axial vector heavy charmed and bottom mesons
 at finite density of the nuclear matter using the chiral SU(3) model
 and QCD sum rules \cite{arv1}. In this approach
 we evaluated the medium modifications of quark and gluon condensates 
 in the nuclear matter through the 
 medium modification of scalar isoscalar fields
 $\sigma$ and $\zeta$ and the scalar dilaton field $\chi$.
In the present paper our objective is to work out the
 in-medium masses and decay constants 
 of heavy charmed vector  
 $\left( D^{\ast}, D_{s}^{\ast} \right)$ 
and axial-vector  $\left( D_{1}, D_{1s}\right)$ 
as well as bottom vector $\left(B^{\ast}, B_{s}^{\ast}\right)$
and axial-vector $\left( B_{1}, B_{1s}\right)$ mesons
 in the isospin asymmetric strange hadronic medium using the
chiral SU(3) model and QCD sum rules. 
In the QCD sum rules the properties of above mesons
are modified through the quark and gluon condensates.
Within the chiral hadronic model the
 quark and gluon condensates are written 
 in terms of scalar fields
$\sigma$, $\zeta$ and $\delta$ and the
scalar dilaton field $\chi$. We shall evaluate the $\sigma$, 
$\zeta$, $\delta$ and $\chi$ fields
and hence the quark and gluon condensates 
in the medium consisting of nucleons and hyperons.
We shall evaluate the values of quark and gluon condensates
as a function of density of strange hadronic medium for different
strangeness fractions and shall find the 
mass shift and shift of decay constants of heavy vector
 and axial vector mesons.
The study of decay constant of heavy mesons
 play important role in understanding the
 strong decay of heavy mesons, their electromagnetic
 structure as well radiative decay width.
The study of B meson decay constants
is important for $B_d$-$\bar{B}_d$
and $B_s$-$\bar{B}_s$ 
mixing \cite{ebert1}.
The  decay constants of vector $D^{*}$
and $B^{*}$ mesons 
are helpful for calculations of strong coupling
in $D^{*}D\pi$ and $B^{*}B\pi$
mesons using light cone sum rules \cite{bel1}.
An extensive literature is available on
the calculations of decay constants
of heavy mesons in the free space,
for example, the QCD sum
rules based on operator product expansion of
two-point correlation function,
heavy-quark expansion \cite{shurya1}, sum rules
in heavy-quark effective theory \cite{heq1,heq2} and
sum rules with gluon radiative corrections
to the correlation functions upto
two loop \cite{gel2,gel3,gel4} or three loop \cite{gel1}.
However, the in-medium modifications
of  decay constants
of heavy mesons had been studied
very recently  in symmetric nuclear matter
only \cite{azzi,wang3}.
  The thermal modification of 
 decay constants of heavy vector mesons 
 was investigated in ref. \cite{azzi2}
 and it was observed  that the values of decay constants remained almost
 constant upto 100 MeV but above this
 decrease sharply with 
 increase in temperature.
  In the present work we shall 
include the contribution of hyperons
in addition to nucleons for evaluating the modification
of mesons properties in asymmetric matter.
  
We shall present this work as follows:
In section (II) we shall describe the chiral SU(3) model 
which is used to evaluate the quark and gluon condensates 
in the strange hadronic matter. Section (III) will introduce 
the QCD sum rules which we shall use in the present 
work along with chiral model to evaluate the
in-medium properties of mesons. In section (IV) we shall present
our results of present investigation and possible discussion
on these results. Section (V) will summarize the present work.

 \section{Chiral SU(3) model}
In this section we shall discuss the chiral SU(3) model
to be used in the present work for evaluation of 
quark and gluon condensates in the strange hadronic medium.
The basic theory of strong interaction, the QCD, is not 
directly applicable in the non-perturbative regime.
To overcome this limitation the effective theories are
constructed which are constrained by the basic 
properties like chiral symmetry and scale invariance of
QCD. The chiral SU(3) model is one such effective model
 based on the non-linear realization
 and broken scale invariance as well as
 spontaneous breaking  properties of chiral symmetry.
 The glueball field $\chi$ is introduced in the model
  to account for the broken scale invariance properties of QCD.
  The model
 had been used successfully in the literature  to study the
 properties of hadrons at finite density and temperature of the
 nuclear and strange hadronic  medium. The 
 general Lagrangian density of the chiral SU(3) model 
 involve the kinetic energy terms, the baryon meson interactions, self
 interaction of vector mesons, scalar mesons-meson 
 interactions as well as the explicit
 chiral symmetry breaking term and is written as
 
 \begin{equation}
{\cal L} = {\cal L}_{kin}+\sum_{W=X,Y,V,A,u} {\cal L}_{BW} + 
{\cal L}_{vec} + {\cal L}_{0} + {\cal L}_{SB}
\label{genlag}
\end{equation}
The details of above Lagrangian density can be found in the reference
\cite{paper3}.

 From the Lagrangian densities
  of the chiral SU(3) model, using
the mean field approximation, we find the coupled equations of motion
for the scalar fields, $\sigma$, $\zeta$, $\delta$ and  the
scalar dilaton field $\chi$ in the isospin asymmetric
strange hadronic medium and  these are

\begin{eqnarray}
&& k_{0}\chi^{2}\sigma-4k_{1}\left( \sigma^{2}+\zeta^{2}
+\delta^{2}\right)\sigma-2k_{2}\left( \sigma^{3}+3\sigma\delta^{2}\right)
-2k_{3}\chi\sigma\zeta \nonumber\\
&-&\frac{d}{3} \chi^{4} \bigg (\frac{2\sigma}{\sigma^{2}-\delta^{2}}\bigg )
+\left( \frac{\chi}{\chi_{0}}\right) ^{2}m_{\pi}^{2}f_{\pi}
-\sum g_{\sigma i}\rho_{i}^{s} = 0 
\label{sigma}
\end{eqnarray}
\begin{eqnarray}
&& k_{0}\chi^{2}\zeta-4k_{1}\left( \sigma^{2}+\zeta^{2}+\delta^{2}\right)
\zeta-4k_{2}\zeta^{3}-k_{3}\chi\left( \sigma^{2}-\delta^{2}\right)\nonumber\\
&-&\frac{d}{3}\frac{\chi^{4}}{\zeta}+\left(\frac{\chi}{\chi_{0}} \right) 
^{2}\left[ \sqrt{2}m_{k}^{2}f_{k}-\frac{1}{\sqrt{2}} m_{\pi}^{2}f_{\pi}\right]
 -\sum g_{\zeta i}\rho_{i}^{s} = 0 
\label{zeta}
\end{eqnarray}
\begin{eqnarray}
&& k_{0}\chi^{2}\delta-4k_{1}\left( \sigma^{2}+\zeta^{2}+\delta^{2}\right)
\delta-2k_{2}\left( \delta^{3}+3\sigma^{2}\delta\right) +2k_{3}\chi\delta 
\zeta \nonumber\\
& + &  \frac{2}{3} d \chi^4 \left( \frac{\delta}{\sigma^{2}-\delta^{2}}\right)
-\sum g_{\delta i}\rho_{i}^{s} = 0
\label{delta}
\end{eqnarray}
\begin{eqnarray}
&& k_{0}\chi \left( \sigma^{2}+\zeta^{2}+\delta^{2}\right)-k_{3}
\left( \sigma^{2}-\delta^{2}\right)\zeta + \chi^{3}\left[1
+{\rm {ln}}\left( \frac{\chi^{4}}{\chi_{0}^{4}}\right)  \right]
+(4k_{4}-d)\chi^{3}
\nonumber\\
& - & \frac{4}{3} d \chi^{3} {\rm {ln}} \Bigg ( \bigg (\frac{\left( \sigma^{2}
-\delta^{2}\right) \zeta}{\sigma_{0}^{2}\zeta_{0}} \bigg ) 
\bigg (\frac{\chi}{\chi_0}\bigg)^3 \Bigg ) \nonumber\\
& + & \frac{2\chi}{\chi_{0}^{2}}\left[ m_{\pi}^{2}
f_{\pi}\sigma +\left(\sqrt{2}m_{k}^{2}f_{k}-\frac{1}{\sqrt{2}}
m_{\pi}^{2}f_{\pi} \right) \zeta\right]  = 0 
\label{chi}
\end{eqnarray}
respectively. 
The values of parameter $k_0$, $k_1$, $k_2$, $k_3$,
$k_4$ and $d$
appearing in above equations are
$2.54$, $1.35$, $-4.78$, $-2.77$, $0.22$  and $0.064$ respectively.
These parameters of the model are fitted so as to ensure 
extrema in the vacuum for the $\sigma$, $\zeta$ and $\chi$ field 
equations, to  reproduce the vacuum masses of the $\eta$ and $\eta '$ 
mesons, the mass of the $\sigma$ meson around 500 MeV, and,
pressure, p($\rho_0$)=0,
with $\rho_0$ as the nuclear matter saturation density \cite{paper3,amarind}.
The values of pion decay constant, $f_{\pi}$ and kaon decay constant, $f_{K}$ are
$93.3$ and $122$ MeV respectively.
The vacuum values of the scalar isoscalar fields, $\sigma$ and $\zeta$ 
and the dilaton field $\chi$ are $-93.3$ MeV, $-106.6$ MeV and 409.8 MeV
respectively. 
The values, $g_{\sigma N} = 10.6$ and $g_{\zeta N} = -0.47$ are 
determined by fitting to the vacuum baryon masses. The other parameters 
fitted to the asymmetric nuclear matter saturation properties 
in the mean-field approximation are: $g_{\omega N}$ = 13.3, 
$g_{\rho p}$ = 5.5, $g_{4}$ = 79.7, $g_{\delta p}$ = 2.5, 
$m_{\zeta}$ = 1024.5 MeV, $ m_{\sigma}$ = 466.5 MeV 
and $m_{\delta}$ = 899.5 MeV. 
In equations (\ref{sigma}) to (\ref{chi}), ${\rho_i}^s$ denote the scalar densities for the baryons 
and at zero temperature are given by expression, 
\begin{eqnarray}
\rho_{i}^{s} = \gamma_{i}\int\frac{d^{3}k}{(2\pi)^{3}} 
\frac{m_{i}^{*}}{E_{i}^{*}(k)}, 
\label{scaldens}
\end{eqnarray}
where, ${E_i}^*(k)=(k^2+{{m_i}^*}^2)^{1/2}$, and, ${\mu _i}^* 
=\mu_i -g_{\omega i}\omega -g_{\rho i}\rho -g_{\phi i}\phi$, are the single 
particle energy and the effective chemical potential
for the baryon of species $i$, and,
$\gamma_i$=2 is the spin degeneracy factor \cite{isoamss}.

In the present work, for the evaluation of vector and axial vector
meson properties using QCD sum rules, we shall need the
light quark condensates $\langle\bar{u}u\rangle$ and $\langle\bar{d}d\rangle$, 
the strange quark condensate $\langle\bar{s}s\rangle$ and the
scalar gluon condensate $\langle \frac{\alpha_s}{\pi}G_{\mu\nu}^{a} G^{\mu\nu a}\rangle$. 
In the chiral effective model the explicit symmetry breaking term
 is introduced to
eliminate the Goldstone bosons
  and can be
used to extract the scalar quark condensates, $\langle q\bar{q}\rangle$ 
in terms of scalar fields $\sigma$, $\zeta$, $\delta$ and $\chi$.
We write 
 \begin{eqnarray}
&&\sum_i m_i \bar {q_i} q_i = - {\cal L} _{SB} \\
& =& \left( \frac {\chi}{\chi_{0}}\right)^{2} 
\left( \frac{1}{2} m_{\pi}^{2} 
f_{\pi} \left( \sigma + \delta \right) +
\frac{1}{2} m_{\pi}^{2} 
f_{\pi} \left( \sigma - \delta \right)
 + \big( \sqrt {2} m_{k}^{2}f_{k} - \frac {1}{\sqrt {2}} 
m_{\pi}^{2} f_{\pi} \big) \zeta \right). 
\label{quark_expl}
\end{eqnarray}
 From equation (\ref{quark_expl}) light  
 scalar quark condensates $\left\langle\bar{u}u\right\rangle $,
  $\left\langle\bar{d}d\right\rangle$ and strange quark
 condensate $\left\langle  \bar{s}s\right\rangle $
  can be written as
 \begin{equation}
\left\langle \bar{u}u\right\rangle 
= \frac{1}{m_{u}}\left( \frac {\chi}{\chi_{0}}\right)^{2} 
\left[ \frac{1}{2} m_{\pi}^{2} 
f_{\pi} \left( \sigma + \delta \right) \right],
\label{qu}
\end{equation}
\begin{equation}
\left\langle \bar{d}d\right\rangle 
= \frac{1}{m_{d}}\left( \frac {\chi}{\chi_{0}}\right)^{2} 
\left[ \frac{1}{2} m_{\pi}^{2} 
f_{\pi} \left( \sigma - \delta \right) \right],
\label{qd}
\end{equation}
and
\begin{equation}
\left\langle \bar{s}s\right\rangle 
= \frac{1}{m_{s}}\left( \frac {\chi}{\chi_{0}}\right)^{2} 
\left[ \big( \sqrt {2} m_{k}^{2}f_{k} - \frac {1}{\sqrt {2}} 
m_{\pi}^{2} f_{\pi} \big) \zeta \right],
\label{qs}
\end{equation}
respectively.

As said earlier the scalar gluon condensates can be 
evaluated in the chiral effective model
through the scalar dilaton field $\chi$.
The broken scale invariance property of QCD implies
that the trace of energy momentum tensor is
non zero (trace anomaly) and is equal to the scalar gluon condensates
for massless QCD i.e.
\begin{equation}
T_{\mu}^{\mu} = \frac{\beta_{QCD}}{2g} 
{G^a}_{\mu\nu} G^{\mu\nu a}
\label{trace_gluonqcd}
\end{equation}
We shall try to find the above trace of energy momentum tensor
within chiral effective model.
The trace anomaly  property 
 of QCD can be mimicked  in the chiral effective model
 through the scale breaking Lagrangian density, 
 \begin{eqnarray}
{\cal L}_{\rm {scalebreaking}} & = &  -\frac{1}{4} \chi^{4} {\rm {ln}}
\Bigg ( \frac{\chi^{4}} {\chi_{0}^{4}} \Bigg )
+ \frac{d}{3}{\chi ^4} 
{\rm {ln}} \Bigg ( \bigg (\frac{I_{3}}{{\rm {det}}\langle X 
\rangle _0} \bigg ) \bigg ( \frac {\chi}{\chi_0}\bigg)^3 \Bigg ),
\label{scalebreak}
\end{eqnarray}
where $I_3={\rm {det}}\langle X \rangle$, with $X$ as the multiplet
for the scalar mesons.

We write the energy momentum tensor for the dilaton field as,
\begin{eqnarray}
T_{\mu \nu}=(\partial _\mu \chi) 
\Bigg (\frac {\partial {{\cal L}_\chi}}
{\partial (\partial ^\nu \chi)}\Bigg )
- g_{\mu \nu} {\cal L}_\chi,
\label{energymom}
\end{eqnarray}
where the Lagrangian density for the dilaton field is,
\begin{eqnarray}
{\cal L}_\chi & = & \frac {1}{2} (\partial _\mu \chi)(\partial ^\mu \chi)
- k_4 \chi^4 \nonumber \\ & - & \frac{1}{4} \chi^{4} {\rm {ln}} 
\Bigg ( \frac{\chi^{4}} {\chi_{0}^{4}} \Bigg )
+ \frac {d}{3} \chi^{4} {\rm {ln}} \Bigg (\bigg( \frac {\left( \sigma^{2} 
- \delta^{2}\right) \zeta }{\sigma_{0}^{2} \zeta_{0}} \bigg) 
\bigg (\frac {\chi}{\chi_0}\bigg)^3 \Bigg ),
\label{lagchi}
\end{eqnarray}
Multiplying 
equation (\ref{energymom}) by $g^{\mu \nu}$, we obtain the trace of 
the energy momentum tensor within the chiral SU(3) model as
\begin{equation}
T_{\mu}^{\mu} = (\partial _\mu \chi) \Bigg (\frac {\partial {{\cal L}_\chi}}
{\partial (\partial _\mu \chi)}\Bigg ) -4 {{\cal L}_\chi}. 
\label{tensor}
\end{equation}
Using the Euler-Lagrange's equation for the $\chi$ field,
the trace of the energy momentum tensor in the chiral SU(3) model
can be expressed as \cite{amarvind,heide1}
\begin{equation}
T_{\mu}^{\mu} = \chi \frac{\partial {{\cal L}_\chi}}{\partial \chi} 
- 4{{\cal L}_\chi} = -(1-d)\chi^{4}.
\label{tensor1}
\end{equation}
Comparing equations (\ref{trace_gluonqcd}) and (\ref{tensor1}), we get the following relation between 
the scalar gluon condensates and the scalar dilaton field $\chi$ (in massless QCD),
\begin{equation}
\frac{\beta_{QCD}}{2g} 
{G^a}_{\mu\nu} G^{\mu\nu a} = -(1-d)\chi^{4}.
\label{gluon_chi}
\end{equation} 
In the case of finite quark masses, the trace of energy momentum tensor is 
written as,
\begin{equation}
T_{\mu}^{\mu} = \sum_i m_i \bar {q_i} q_i+ \langle \frac{\beta_{QCD}}{2g} 
G_{\mu\nu}^{a} G^{\mu\nu a} \rangle  \equiv  -(1 - d)\chi^{4}, 
\label{tensor2m}
\end{equation}
where the first term of the energy-momentum tensor, within the chiral 
SU(3) model is the negative of the explicit chiral symmetry breaking
term, ${\cal L}_{SB}$ (see equation (\ref{quark_expl})).

The QCD $\beta$ function at one loop level, for 
$N_{c}$ colors and $N_{f}$ flavors is given by
\begin{equation}
\beta_{\rm {QCD}} \left( g \right) = -\frac{11 N_{c} g^{3}}{48 \pi^{2}} 
\left( 1 - \frac{2 N_{f}}{11 N_{c}} \right)  +  O(g^{5})
\label{beta}
\end{equation}
In the above equation, the first term in the parentheses arises from 
the (antiscreening) self-interaction of the gluons and the second term, 
proportional to $N_{f}$, arises from the (screening) contribution of 
quark pairs.

The trace of the energy-momentum tensor in QCD, using the 
one loop beta function given by equation (\ref{beta}),
for $N_c$=3 and $N_f$=3, and accounting for the finite quark masses
\cite{cohen} is given as,
\begin{equation}
T_{\mu}^{\mu} = - \frac{9}{8} \frac{\alpha_{s}}{\pi} 
{G^a}_{\mu\nu} {G^a}^{\mu\nu}
+\left( \frac {\chi}{\chi_{0}}\right)^{2} 
\left( m_{\pi}^{2} 
f_{\pi} \sigma
+ \big( \sqrt {2} m_{k}^{2}f_{k} - \frac {1}{\sqrt {2}} 
m_{\pi}^{2} f_{\pi} \big) \zeta \right). 
\label{tensor4}
\end{equation} 
Using equations (\ref{tensor}) and (\ref{tensor4}), we can write  
\begin{equation}
\left\langle  \frac{\alpha_{s}}{\pi} {G^a}_{\mu\nu} {G^a}^{\mu\nu} 
\right\rangle =  \frac{8}{9} \Bigg [(1 - d) \chi^{4}
+\left( \frac {\chi}{\chi_{0}}\right)^{2} 
\left( m_{\pi}^{2} f_{\pi} \sigma
+ \big( \sqrt {2} m_{k}^{2}f_{k} - \frac {1}{\sqrt {2}} 
m_{\pi}^{2} f_{\pi} \big) \zeta \right) \Bigg ]. 
\label{chiglu}
\end{equation}
We thus see from the equation (\ref{chiglu}) that the scalar 
gluon condensate $\left\langle \frac{\alpha_{s}}{\pi} G_{\mu\nu}^{a} 
G^{\mu\nu a}\right\rangle$ is related to the dilaton field $\chi$.
For massless quarks, since the second term in (\ref{chiglu}) 
arising from explicit symmetry breaking is absent, the scalar 
gluon condensate becomes proportional to the fourth power of the 
dilaton field, $\chi$, in the chiral SU(3) model.

 \section{QCD sum rules for vector and axial-vector heavy mesons in strange
 hadronic matter}
 In this section we shall discuss the QCD sum rules \cite{wang1,wang2} which will be
used later along with the chiral SU(3) model for the evaluation of 
in-medium properties of vector and axial vector mesons
in asymmetric strange hadronic matter. 
To find the mass modification of above discussed heavy mesons we shall
 use the two-point correlation function $\Pi_{\mu\nu}(q)$,
 \begin{eqnarray}
\Pi_{\mu\nu}(q) &=& i\int d^{4}x\ e^{iq \cdot x} \langle T\left\{J_\mu(x)J_\nu^{\dag}(0)\right\} \rangle_{\rho_B} \,.
 \end{eqnarray}
 In above equation $J_\mu(x)$ denotes the isospin 
 averaged current, $x = x^\mu = (x^0,\textbf{x})$ is the four coordinate,
  $q = q^\mu = (q^0,\textbf{q})$ is four momentum and $T$ denotes the
 time ordered operation on the product of quantities in the brackets. From above definition it is clear
 that the two point correlation function is actually a Fourier transform 
 of the expectation value of the
 time ordered product of two currents.
 
 For the vector and axial vector mesons  average particle-antiparticle currents are given by the
 expressions

\begin{eqnarray}
 J_\mu(x) &= &J_\mu^\dag(x) =\frac{\bar{c}(x)\gamma_\mu q(x)+\bar{q}(x)\gamma_\mu c(x)}{2}\, , \nonumber\\
 \label{vectorcurrent}
 \end{eqnarray}
 and
 \begin{eqnarray}
  J_{5\mu}(x) &=&J_{5\mu}^\dag(x) =\frac{\bar{c}(x)\gamma_\mu \gamma_5q(x)+\bar{q}(x)\gamma_\mu\gamma_5 c(x)}{2}\,,
  \label{axialcurrent}
\end{eqnarray}
respectively.
 Note that in above equations $q$ denotes the  $u$, $d$ or $s$ quark
 (depending upon type of  meson under investigation) whereas $c$ denotes the heavy charm quark
 (for B mesons c quark will be replaced by bottom b quark).
Also note that in the present work, instead of considering the mass splitting 
between particles and antiparticles, 
we emphasize on the mass shift and mass-splitting of isospin doublet 
$D$ $\left( D^{+}, D^{0}\right) $ and
 $B$ $\left( B^{+}, B^{0}\right)$ mesons corresponding to
 both vector and axial vector mesons.
To find the mass splitting of particles
and antiparticles in the nuclear medium one has to consider the even and 
odd part of QCD sum rules \cite{higler1}.
For example, in ref. \cite{higler1} the mass splitting between 
pseudoscalar $D$
and $\bar{D}$ mesons was investigated using the even and odd QCD sum rules
whereas in \cite{haya1,wang1,wang2} the mass-shift of $D$ mesons was investigated
under centroid approximation.   
  
 At zero temperature the two point correlation function can be decomposed into
  the vacuum part and a static one-nucleon part 
   i.e. we can write
 \begin{eqnarray}
\Pi_{\mu\nu}(q) &=&\Pi^{0}_{\mu\nu}(q)+ \frac{\rho_B}{2M_N}T^{N}_{\mu\nu}(q),
\label{pibn}
 \end{eqnarray}
where
\begin{eqnarray}
T^{N}_{\mu\nu}(\omega,\mbox{\boldmath $q$}\,) &=&i\int d^{4}x e^{iq\cdot x}\langle N(p)|
T\left\{J_\mu(x)J_\nu^{\dag}(0)\right\} |N(p) \rangle\,.
\end{eqnarray}
In above equation $|N(p)\rangle$ denotes the isospin and spin averaged static nucleon state
with the four-momentum $p = (M_{N},0)$. The state is normalized as 
 $\langle N(\mbox{\boldmath $p$})|N(\mbox{\boldmath
$p$}')\rangle = (2\pi)^{3} 2p_{0}\delta^{3}(\mbox{\boldmath
$p$}-\mbox{\boldmath $p$}')$.

As discussed in Ref. \cite{wang2}, in the limit of the $3$-vector  $\mbox{\boldmath $q$}\rightarrow {\bf 0}$, the
correlation functions $T_{N}(\omega,\mbox{\boldmath $q$}\,)$ can be related to the $D^{*}N$ and $D_{1}N$ 
 scattering T-matrices. Thus we write \cite{wang2}
 \begin{eqnarray}
 {{\cal T}_{D^*N}}(M_{D^*},0)= 8\pi(M_N+M_{D^*})a_{D^*} \nonumber\\
{{\cal T}_{D_1N}}(M_{D_1},0) = 8\pi(M_N+M_{D_1})a_{D_1}
\end{eqnarray}
In above equation $a_{D^*}$ and $a_{D_1}$ are the scattering lengths of $D^*N$ and $D_1N$
respectively.
Near the pole positions of vector and axial vector mesons
the phenomenological spectral densities can be parametrized with three unknown parameters
$a, b$ and $c$ i.e. we write \cite{wang1,wang2,haya1}
\begin{eqnarray}
\rho(\omega,0) &=& -\frac{f_{D^*/D_1}^2M_{D^*/D_1}^2}{\pi}
 \mbox{Im} \left[\frac{{{\cal T}_{D^*/D_1N}}(\omega,{\bf 0})}{(\omega^{2}-
M_{D^*/D_1}^2+i\varepsilon)^{2}} \right]\nonumber\\
&+& \cdots = a\,\frac{d}{d\omega^2}\delta^{'}(\omega^{2}-M_{D^*/D_1}^2)
 +
b\,\delta(\omega^{2}-M_{D^*/D_1}^2) + c\,\theta(\omega^{2}-s_{0})\,.
\label{a1}
\end{eqnarray}
The term denoted by $...$ represent the continuum contributions.
The first term denotes the double-pole term and corresponds to 
the on-shell effects of the T-matrices,

\begin{eqnarray}
a=-8\pi(M_N+M_{D^*/D_1})
 a_{D^*/D_1}f_{D^*/D_1}^2M_{D^*/D_1}^2\,.
\label{a2}
\end{eqnarray}

Now we shall write the relation between the scattering length of mesons and
their in-medium mass-shift. For this first we note that
the shift of squared mass of mesons can be written in 
terms of the parameter $a$ appearing in equation (\ref{a1}) through relation \cite{koike1},
\begin{eqnarray}
&&\Delta m_{D^*/D_1}^{2} = \frac{\rho_B }{2M_N} \frac{a }{f_{D^*/D_1}^2M_{D^*/D_1}^2} \nonumber\\
&=& -\frac{\rho_B }{2M_N}8\pi(M_N+M_{D^*/D_1})a_{D^*/D_1}\,, 
\label{massshiftsq}
\end{eqnarray}
where in the last term we used equation (\ref{a2}).
The mass shift is now defined by the relation
\begin{eqnarray}
\delta m_{D^*/D_1} = \sqrt{m_{D^*/D_1}^2+\Delta m_{D^*/D_1}^{2}}
- m_{D^*/D_1} .
\label{massshift}
\end{eqnarray}
The second term in equation (\ref{a1}) denotes  the single-pole term,
and corresponds to the off-shell (i.e. $\omega^2\neq M_{D^*/D_1}^2$) effects of the $T$-matrices. The
third term denotes   the continuum term or the  remaining effects,
where, $s_{0}$, is the continuum threshold.
The continuum threshold parameter $s_0$ define the scale below which
the continuum contribution
vanishes \cite{kwon1}. 

The shift in the decay constant of vector or axial vector mesons 
can be written as \cite{wang3},
\begin{eqnarray}
\delta f_{D^{\star}/ D_1} =  \frac{1}{2f_{D^{\star}/D_1}m_{D^{\star}/D_1}^{2}}
\left(\frac{\rho_{B}}{2m_{N}}b - 
2f_{D^{\star}/D_1}^{2} m_{D^{\star}/D_1}
\delta m_{D^{\star}/D_1} \right) .
\label{decayshift}
\end{eqnarray}

From  equations
 (\ref{massshift}) and (\ref{decayshift}) we observe 
 that  to find the value of mass shift and shift in decay constant
 of mesons we first need to find the value of unknown parameters $a$ and $b$.
These can be determined as follows: we note that in the low energy limit, 
$\omega\rightarrow 0$, the
$T_{N}(\omega,{\bf 0})$ is equivalent to the Born term $T_{D^*/D_1N}^{\rm
Born}(\omega,{\bf 0})$. We take
into account the Born term at the phenomenological side,
\begin{eqnarray}
T_{N}(\omega^2)&=&T_{D^*/D_1N}^{\rm
Born}(\omega^2)+\frac{a}{(M_{D^*/D_1}^2-\omega^2)^2}\nonumber\\
&+&\frac{b}{M_{D^*/D_1}^2-\omega^2}+\frac{c}{s_0-\omega^2}\, ,
\label{bornconst}
\end{eqnarray}
with the constraint
\begin{eqnarray}
\frac{a}{M_{D^*/D_1}^4}+\frac{b}{M_{D^*/D_1}^2}+\frac{c}{s_0}&=&0 \, .
\label{constraint}
\label{aconst}
\end{eqnarray}
Note that in Eq. (\ref{bornconst}) the phenomenological side of 
scattering amplitude for $q_{\mu}\neq 0$ is
not exactly equal to Born term but there are contributions from
other terms. However, for $\omega\rightarrow 0$, $T_N$ on left should
be equal to $T^{Born}$ on right side of Eq. (\ref{bornconst})
and this requirement results in constraint given in Eq. (\ref{constraint}).
As we shall discuss below the constraint
 (\ref{constraint}) help in eliminating the parameter $c$
 and scattering amplitude will be function of parameters $a$ and $b$ only.
The Born terms to be used in equation (\ref{bornconst}) for
  vector and axial-vector mesons are given by
following relations \cite{wang1,wang2}
 \begin{eqnarray}		
T_{D^*N}^{\rm Born}(\omega,{\bf0})&=&\frac{2f_{D^*}^2M_{D^*}^2M_N(M_H+M_N)g_{D^*NH}^2}
{\left[\omega^2-(M_H+M_N)^2\right]\left[\omega^2-M_{D^*}^2\right]^2}\,, \nonumber \\
T_{D_1N}^{\rm Born}(\omega,{\bf0})&=&\frac{2f_{D_1}^2M_{D_1}^2M_N(M_H-M_N)g_{D_1NH}^2}
{\left[\omega^2-(M_H-M_N)^2\right]\left[\omega^2-M_{D_1}^2\right]^2}\,.
\end{eqnarray}
In the above equations 
$g_{D^{\star}NH}$ and $g_{D_1NH}$ are the coupling constants.
 $M_H$ is the the mass of the hadron e.g. corresponding to charm mesons
 we have $\Lambda_c$ and $\Sigma_c$ whereas corresponding to bottom mesons
 we have the hadrons $\Lambda_b$ and $\Sigma_b$.
Corresponding to charm mesons we take the average value of the masses of $M_{\Lambda_c}$ and $M_{\Sigma_c}$ and 
 is equal to $2.4$ GeV. For the case of mesons having bottom quark, $b$, we
 consider the average value of masses 
 of $\Lambda_b$ and $\Sigma_b$ and it is equal to 5.7 GeV \cite{wang2,wang3}.

Now we write the equation for the Borel transformation of the scattering matrix on the
phenomenological side and equate that to the Borel transformation of the scattering matrix for the 
operator expansion side. 
For vector meson, $D^{*}$, the Borel transformation equation is given by \cite{wang2},
\begin{eqnarray}
&& a \left\{\frac{1}{M^2}e^{-\frac{M_{D^*}^2}{M^2}}-\frac{s_0}{M_{D^*}^4}e^{-\frac{s_0}{M^2}}\right\} 
+b \left\{e^{-\frac{M_{D^*}^2}{M^2}}-\frac{s_0}{M_{D^*}^2}e^{-\frac{s_0}{M^2}}\right\}\nonumber\\
&+& B  \left[\frac{1}{(M_H+M_N)^2-M_{D^*}^2}-\frac{1}{M^2}\right]
e^{-\frac{M_{D^*}^2}{M^2}}
-\frac{B}{(M_H+M_N)^2-M_{D^*}^2}e^{-\frac{(M_H+M_N)^2}{M^2}}\nonumber\\
&=&\left\{-\frac{m_c\langle\bar{q}q\rangle_N}{2}\right.
\left.-\frac{2\langle q^\dag i D_0q\rangle_N}{3}+\frac{m_c^2\langle q^\dag i D_0q\rangle_N}{M^2}
\right\}e^{-\frac{m_c^2}{M^2}}
+\frac{m_c\langle\bar{q}g_s\sigma Gq\rangle_N}{3M^2}e^{-\frac{m_c^2}{M^2}}\nonumber\\
&+&\left\{\frac{8m_c\langle \bar{q} i D_0 i D_0q\rangle_N}{3M^2}-\frac{m_c^3\langle \bar{q} i D_0 i D_0q\rangle_N}{M^4}\right\}e^{-\frac{m_c^2}{M^2}}\nonumber\\
&&-\frac{1}{24}\langle\frac{\alpha_sGG}{\pi}\rangle_N\int_0^1 dx \left(1+\frac{\widetilde{m}_c^2}{2M^2}\right)e^{-\frac{\widetilde{m}_c^2}{M^2}}\nonumber\\
&+&\frac{1}{48M^2}\langle\frac{\alpha_sGG}{\pi}\rangle_N\int_0^1 dx\frac{1-x}{x}\left(\widetilde{m}_c^2-\frac{\widetilde{m}_c^4}{M^2}\right)e^{-\frac{\widetilde{m}_c^2}{M^2}}\,.
\label{qcdsumdst}
\end{eqnarray}
where, $B = \frac{2f_{D^*}^2M_{D^*}^2M_N(M_H+M_N)g_{D^*NH}^2}{(M_H+M_N)^2-M_{D^*}^2}$.
Note that in equation (\ref{qcdsumdst}) we have two unknown parameters $a$ and $b$.
We differentiate  equation (\ref{qcdsumdst})  w.r.t.
$\frac{1}{M^{2}}$  so that we could have two equations and two unknowns. By solving those two coupled equations we will be able to get the
values of parameters $a$ and $b$. 
Same procedure will be applied to obtain the values of parameters 
$a$ and $b$ corresponding to axial-vector mesons.
For the axial-vector mesons, $D_{1}$, the Borel transformation equation is given by \cite{wang2},
\begin{eqnarray}
&& a \left\{\frac{1}{M^2}e^{-\frac{M_{D_1}^2}{M^2}}-\frac{s_0}{M_{D_1}^4}e^{-\frac{s_0}{M^2}}\right\}
+ b \left\{e^{-\frac{M_{D_1}^2}{M^2}}-\frac{s_0}{M_{D_1}^2}e^{-\frac{s_0}{M^2}}\right\}\nonumber\\
&+& C \left[\frac{1}{(M_H-M_N)^2-M_{D_1}^2}-\frac{1}{M^2}\right]
e^{-\frac{M_{D_1}^2}{M^2}}
-\frac{C}{(M_H-M_N)^2-M_{D_1}^2}e^{-\frac{(M_H-M_N)^2}{M^2}}\nonumber\\
&=&\left\{\frac{m_c\langle\bar{q}q\rangle_N}{2}\right.
\left.-\frac{2\langle q^\dag i D_0q\rangle_N}{3}+\frac{m_c^2\langle q^\dag i D_0q\rangle_N}{M^2}\right\}e^{-\frac{m_c^2}{M^2}}
-\frac{m_c\langle\bar{q}g_s\sigma Gq\rangle_N}{3M^2}e^{-\frac{m_c^2}{M^2}}\nonumber\\
&-&\left\{\frac{8m_c\langle \bar{q} i D_0 i D_0q\rangle_N}{3M^2}-\frac{m_c^3\langle \bar{q} i D_0 i D_0q\rangle_N}{M^4}\right\}e^{-\frac{m_c^2}{M^2}}\nonumber\\
&&-\frac{1}{24}\langle\frac{\alpha_sGG}{\pi}\rangle_N\int_0^1 dx \left(1+\frac{\widetilde{m}_c^2}{2M^2}\right)e^{-\frac{\widetilde{m}_c^2}{M^2}}\nonumber\\
&+&\frac{1}{48M^2}\langle\frac{\alpha_sGG}{\pi}\rangle_N\int_0^1 dx\frac{1-x}{x}\left(\widetilde{m}_c^2-\frac{\widetilde{m}_c^4}{M^2}\right)e^{-\frac{\widetilde{m}_c^2}{M^2}}\,,
\label{qcdsumd1}
\end{eqnarray}
where, $C = \frac{2f_{D_1}^2M_{D_1}^2M_N(M_H-M_N)g_{D_1NH}^2}{(M_H-M_N)^2-M_{D_1}^2}$.
In the above equations  $\widetilde{m}_c^2=\frac{m_c^2}{x}$.

As discussed earlier, in determining the properties of hadrons from QCD sum rules, we shall use the 
values of quark and gluon condensates as calculated using chiral SU(3) model.
Any operator $\mathcal{O}$ on OPE side can be written as \cite{koike1,zscho1,kwon1},
\begin{eqnarray}
\mathcal{O}_{\rho_{B}} &=&\mathcal{O}_{vacuum} +
4\int\frac{d^{3}p}{(2\pi)^{3} 2 E_{p}}n_{F}\left\langle N(p)\vert \mathcal{O}\vert N(p) \right\rangle
 \nonumber\\
& =& \mathcal{O}_{vacuum} + \frac{\rho_B}{2M_N}\mathcal{O}_{N}
\label{operator1}
\end{eqnarray}
In above equation, $\mathcal{O}_{\rho_{B}}$ , gives
us the expectation value of the operator at finite baryonic density. 
The term $\mathcal{O}_{vacuum}$ stands for the vacuum expectation value of the
operator and  $\mathcal{O}_{N}$ gives us the nucleon expectation value
of the operator. 
Thus within chiral SU(3) model, we can find the values of
 $\mathcal{O}_{\rho_{B}}$ at finite density of the nuclear medium and hence can find
 $\mathcal{O}_{N}$ using
\begin{equation}
\mathcal{O}_{N} = \left[ \mathcal{O}_{\rho_{B}}  - \mathcal{O}_{vacuum}\right] \frac{2M_N}{\rho_B}.
\label{condexp}
\end{equation}

The scalar quark condensates  $\left\langle \bar{q}q\right\rangle $
in equations (\ref{qcdsumdst}) and (\ref{qcdsumd1})  are
evaluated using equations (\ref{qu}), (\ref{qd}) and (\ref{qs}).
The condensate $\langle\bar{q}g_s\sigma Gq\rangle_{\rho_B}$ is given by the equation \cite{qcdThomas},
\begin{equation}
\langle\bar{q}g_s\sigma Gq\rangle_{\rho_B} = \lambda^{2}\left\langle \bar{q}q \right\rangle_{\rho_{B}} + 3.0 GeV^{2}\rho_{B}.
\label{quarkcond2}
\end{equation}
Also we write \cite{qcdThomas},
\begin{equation}
\langle \bar{q} i D_0 i D_0q\rangle_{\rho_B} + \frac{1}{8}\langle\bar{q}g_s\sigma Gq\rangle_{\rho_B} =  0.3 GeV^{2}\rho_{B}.
\label{quarkcond3}
\end{equation}
As discussed above the quark condensate, $ \left\langle \bar{q}q\right\rangle _{\rho_{B}} $,
can be calculated within the chiral SU(3) model.
 This value of $ \left\langle \bar{q}q\right\rangle _{\rho_{B}} $
can be used through equations (\ref{quarkcond2}) and (\ref{quarkcond3}) to
calculate the value of condensates $\langle\bar{q}g_s\sigma Gq\rangle_{\rho_B}$ and
 $\langle \bar{q} i D_0 i D_0q\rangle_{\rho_B}$ within chiral SU(3) model.
 The value of quark condensate $\langle q^\dag i D_0q\rangle$ for light quark is equal to  $0.18 GeV^2 \rho_{B}$ and for strange quark it is $0.018 GeV^2 \rho_B$ \cite{qcdThomas}.
 
 It may be noted that the QCD sum rules for the evaluation of in-medium properties of
  vector mesons, $B^*$ and axial vector mesons $B_1$, can be written by replacing 
  masses of charmed mesons  $D^*$ and $D_1$, by  
  corresponding masses of  bottom mesons  $B^*$ and $B_1$ 
  in equations  (\ref{qcdsumdst}) and (\ref{qcdsumd1}) respectively. 
   Also the bare charm quark mass, $m_c$ will be replaced by the mass of bottom quark, $m_b$.

 \section{Results and Discussions}
In this section we shall discuss the results of present
investigation of in-medium mass shift and shift in decay 
constant of 
vector  ($D^{\star} (D^{+ \star}, D^{0 \star}, D_{s}^{\star})$ and  
$B^{\star} (B^{+ \star}, B^{0 \star}, B_{s}^{\star})$) and
axial vector ($D_{1} (D^{+}_{1}, D^{0}_{1}, D_{1s})$ and 
$B_{1} (B^{+}_{1}, B^{0}_{1}, B_{1s})$) mesons in
isospin asymmetric strange hadronic matter. 
First we list the values of various parameters
used in the present work on vector and
axial vector mesons. 
Nuclear matter saturation density adopted in 
the present investigation is 0.15 $fm^{-3}$. 
The value of coupling constants 
$g_{{D^*N\Lambda_c}}$ $\approx$ $g_{{D^*N\Sigma_c}}$ $\approx$ $g_{{D_1N\Lambda_c}}$ $\approx$ $g_{{D_1N\Sigma_c}}$ $\approx$ $g_{{B^*N\Lambda_c}}$ $\approx$ $g_{{B^*N\Sigma_c}}$ $\approx$  $g_{{B_1N\Lambda_c}}$ $\approx$ $g_{{B_1N\Sigma_c}}$
 is 3.86 \cite{wang2}.
   The masses of 
   different mesons 
   $M_{D^{*+}}$, $M_{D^{*0}}$, $M_{B^{*+}}$, 
   $M_{B^{*0}}$, $M_{D_1^{+}}$, $M_{D_1^0}$, 
   $M_{B_1^{+}}$, $M_{B_1^0}$, $M_{D_s^*}$, $M_{B_s^*}$, $M_{D_{1s}}$ and $M_{B_{1s}}$
    used in this present investigation are 2.01, 
    2.006, 5.325, 5.325, 2.423, 2.421, 5.721,
     5.723, 2.112, 5.415, 2.459 and 5.828 GeV
      respectively.
The values of decay constants $f_{D^{*}}$, $f_{B^{*}}$, $f_{D_1}$,
  $f_{B_1}$, $f_{D_s^*}$, $f_{B_s^*}$, $f_{D_{1s}}$ and $f_{B_{1s}}$ are
0.270, 0.195, 0.305, 0.255,  1.16*(0.270), 1.16*(0.195), 1.16*(0.305)
  and 1.16*(0.255) GeV respectively.
  The masses of quarks namely up, u, down, d, strange, s,
   charm, c, and bottom, b, used
  in the present work 
  are 0.005, 0.007, 0.105, 1.4 and 2.3 GeV respectively.
   The values of
   threshold parameter $s_0$ for $D^{*}$,
    $B^{*}$, $D_1$, $B_1$, $D_s^*$, $B_s^*$, $D_{1s}$  and  
      $B_{1s}$ mesons  are taken as 6.5, 35, 
      8.5, 39, 7.5, 38, 9.5 and 41
       GeV$^2$ respectively. 
          The various coupling constants and 
          continuum threshold parameters $s_0$ are not subjected to 
          medium modifications. To describe the exact mass(decay) 
          shift of above mesons we have chosen a suitable Borel
           window i.e. the range of squared Borel mass
           parameter, $M^{2}$, 
           within which there is almost no variation in the mass and
            decay constant.  
            In table  (I) 
            we mention the Borel windows
            as observed in the present calculations for the
            mass shift and shift in decay constant of vector and axial vector 
            mesons.
            \begin{table}
            \begin{center}
\begin{tabular}{ |c|c|c|c|c|c|c|c|c| }
\hline
  & $\delta m_{D^{*+}}$ & $\delta m_{D^{*0}}$ &  $\delta m_{D_s^{*}} $ & $\delta m_{B^{*+}}$ & $\delta m_{B^{*0}}$ & $\delta m_{B_s^{*}} $ \\
 \hline
 $M^2$ (GeV$^{2}$)& (4.5 - 6.5) & (4.5 - 6.5) & (5.0 - 7.0) & (30 - 33) & (30 - 33) & (31 - 34) \\  
 \hline \hline

  & $\delta f_{D^{*+}}$ & $\delta f_{D^{*0}}$ &  $\delta f_{D_s^{*}} $ & $\delta f_{B^{*+}}$ & $\delta f_{B^{*0}}$ & $\delta f_{B_s^{*}} $ \\
 \hline
 $M^2$ (GeV$^{2}$) & (3.3 - 4.9) & (3.3 - 4.9) & (3.8 - 5.3) & (26 - 31) & (26 - 31) & (27 - 31) \\  
 \hline \hline \hline \hline

  & $\delta m_{D_1^{+}}$ & $\delta m_{D_1^{0}}$ &  $\delta m_{D_{1s}} $ & $\delta m_{B_1^{+}}$ & $\delta m_{B_1^{0}}$ & $\delta m_{B_{1s}} $ \\
 \hline
 $M^2$ (GeV$^{2})$ & (5.4 - 9.4) & (5.4 - 9.4) & (5.9 - 9.9) & (33 - 38) & (33 - 38) & (36 - 40) \\  
 \hline \hline

  & $\delta f_{D_1^{+}}$ & $\delta f_{D_1^{0}}$ &  $\delta f_{D_{1s}} $ & $\delta f_{B_1^{+}}$ & $\delta f_{B_1^{0}}$ & $\delta f_{B_{1s}} $ \\
 \hline
 $M^2$ (GeV$^{2})$ & (4.2 - 7.2) & (4.2 - 7.2) & (5.0 - 8.0) & (30 - 34) & (30 - 34) & (32 - 36) \\  
 \hline
     
\end{tabular}
\caption{In the above table we mention the Borel windows as observed for mass shift
and shift in decay constant of vector and axial vector mesons.}
\end{center}
\label{parameters}
         \end{table}   
             
\begin{figure}
\includegraphics[width=16cm,height=14cm]{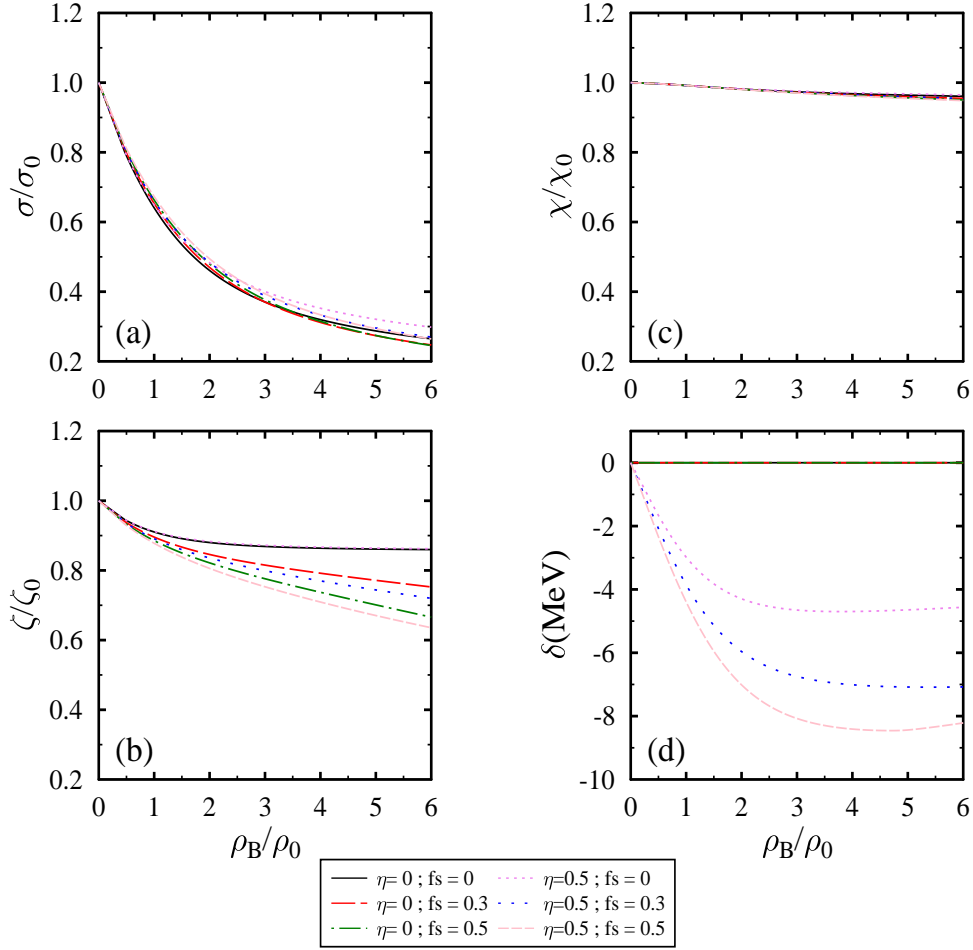}  
\caption{(Color online)
In above figure, subplots (a), (b) and (c) show 
the variation of ratio of in-medium value  to vacuum value
of scalar fields, $\sigma/\sigma_0$, $\zeta/\zeta_0$
and $\chi/\chi_0$
 as a function of
 baryonic density, $ \rho_B$ (in units of nuclear saturation density, $\rho_0$). 
 In subplot
 (d) we have shown the scalar-isovector field $\delta$
 as a function of density of medium. 
 We compare the results at isospin asymmetric parameters $\eta$ = 0 and 0.5. 
 For each value of isospin asymmetry parameter, $\eta$, the results are
 shown for 
 strangeness fractions $f_s$ =0, 0.3 and 0.5.}
\label{fields}
\end{figure}

We start with the discussion on the behaviour of
quark and gluon condensates for different strangeness fractions
 and isospin asymmetry parameters
 of strange hadronic medium. 
In literature, the quark condensates are evaluated to
leading order in nuclear density
using the Feynman Hellmann theorem
and model independent results were obtained 
in terms of pion nucleon sigma term \cite{cohen, quark2}.
Using the Feynman Hellmann theorem, the quark
condensate at finite density of nuclear matter
is expressed as sum of vacuum value and a term dependent
on energy density of nuclear matter.
In model independent calculations
the interactions between nucleons were 
neglected and free space nucleon mass was used.
If one use only the leading order calculations
for the evaluation of
quark condensates above
nuclear matter density
then the quark condensates decreases very sharply 
and almost vanishes around 3$\rho_0$.
In ref. \cite{quark3} Dirac-Brueckner approach
with the Bonn boson-exchange potential
was used to include the higher order corrections
and to find the quark condensates in the
nuclear matter above the
nuclear matter density. The calculations show that
at higher density the quark condensate
decrease more slowly as compared to
leading order predictions.
In the chiral model used in the present work the
quark and gluon condensates are expressed in terms of 
scalar fields $\sigma$, $\zeta$, $\delta$ and
 $\chi$ (see equations (\ref{qu}), (\ref{qd})
(\ref{qs}) and (\ref{chiglu})). 
The scalar fields $\sigma$, $\zeta$, $\delta$ and
 $\chi$ in the strange hadronic
medium are evaluated by solving the
coupled equations (\ref{sigma}), (\ref{zeta}), (\ref{delta}) 
and (\ref{chi}).
In figure (\ref{fields}) we show the
variation of ratio of in-medium value to vacuum value
of scalar fields as a function of baryonic
density of strange hadronic medium. We show
the results for  isospin asymmetry parameter
$\eta$ = 0  and 0.5. For each value of
$\eta$, the results are shown for strangeness
fractions, $f_s$  = 0, 0.3 and 0.5.
From figure one can see that the  scalar field
$\sigma$ and $\zeta$ varies considerably
as a function of baryonic density of
medium whereas the scalar field $\chi$
has little density dependence. 
For example, in symmetric nuclear medium ($\eta$ = 0 and $f_s$ = 0),
at density $\rho_B$ = $\rho_0$ ($4\rho_0$)
the values of scalar fields $\sigma$, $\zeta$ and $\chi$
are observed to be
0.64 $\sigma_0$ (0.31 $\sigma_0$), 
0.91 $\zeta_0$ (0.86 $\zeta_0$), and
0.99 $\chi_0$ (0.97 $\chi_0$). The symbols
$\sigma_0$, $\zeta_0$ and $\chi_0$ denote the vacuum values of
scalar fields and have values
$-93.29$, $-106.75$ and $-409.76$ MeV respectively.
From figure (\ref{fields}) we observe
that at high baryon density the strange scalar-isoscalar field
$\zeta$ varies considerably as a function of
strangeness fraction as compared to 
non-strange scalar isoscalar field $\sigma$.
For example, in symmetric medium ($\eta$ = 0), at baryon density
$4\rho_0$,
as we move from $f_s$ = 0 to $f_s$ = 0.5,
the value of $\zeta$ changes by  14 \%, whereas the
value of $\sigma$ changes by 1\% only.
However, the effect of isospin asymmetry
of the medium is more 
on the values of $\sigma$ field as
compared to $\zeta$ field. For example, in nuclear medium ($f_s$ = 0), at
baryon density $\rho_B$  = 4$\rho_0$,
as we move from $\eta$  = 0 to 0.5 the values of
non-strange scalar field, $\sigma$ and the strange scalar
field, $\zeta$ changes by 10.25\% and 0.24 \% respectively.
However, in strange medium, at $f_s$ = 0.5,
the percentage change in the values of $\sigma$
and $\zeta$ is 7.2\% and 7\% respectively.
Since the scalar meson, $ \sigma$ 
has light quark content ($u$ and $d$ quarks)  and the $\zeta$ meson
have strange quark contents ($s$ quark) and therefore former
is more sensitive to 
isospin asymmetry of the  medium 
(property of $u$ and $d$ quarks) and the
latter is to
strangeness fraction of the medium.

\begin{figure}
\includegraphics[width=16cm,height=16cm]{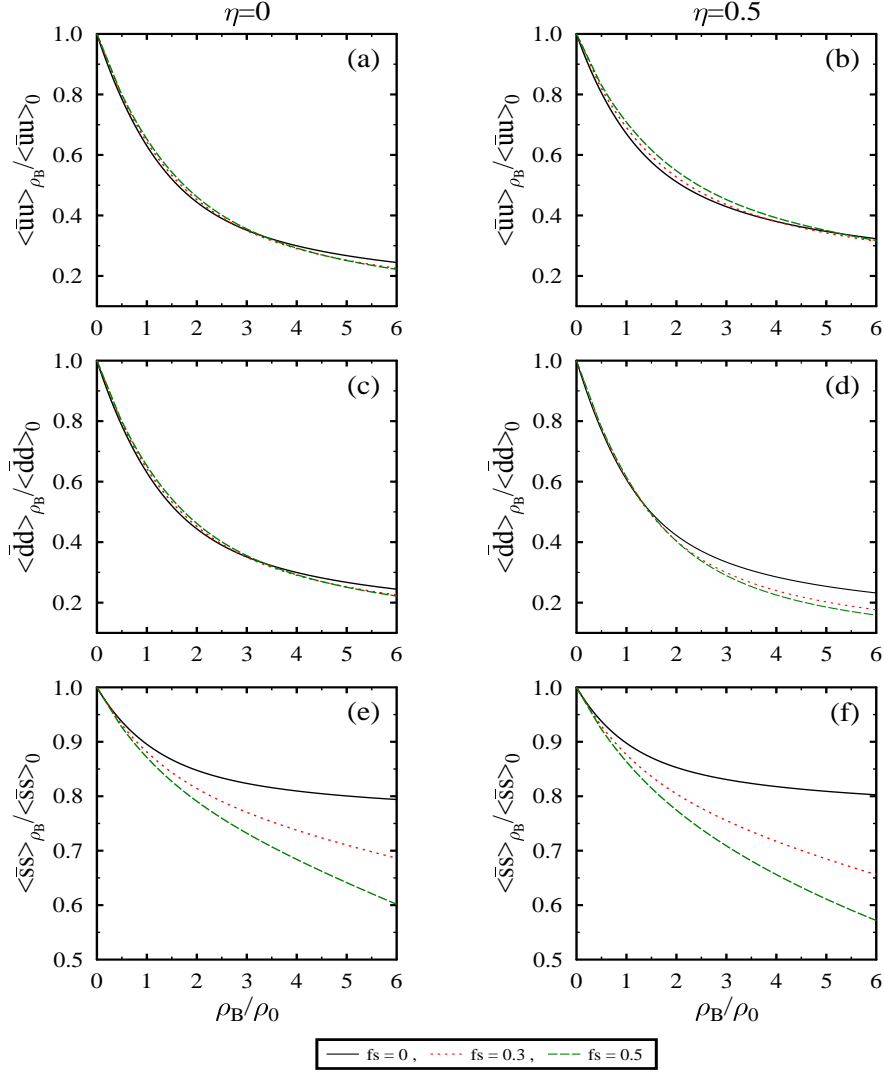}  
\caption{(Color online)
In above figure we plot the ratio of in-medium quark condensate to vacuum condensate
 as a function of
 baryonic density, $ \rho_B$ (in units of nuclear saturation density, $\rho_0$). 
 We compare the results at isospin asymmetric parameters $\eta$ = 0 and 0.5. 
 For each value of isospin asymmetry parameter, $\eta$, the results are
 shown for 
 strangeness fractions $f_s$ =0, 0.3 and 0.5.}
\label{lu}
\end{figure}

In figure (\ref{lu})
we show the variation of  ratio of  in-medium value of 
quark condensate to the vacuum value of condensate
 as a function of baryonic density of hadronic medium
for different values of strangeness 
fractions $f_{s}$ and isospin asymmetry parameter $\eta$.
We show  the results for light quark condensates 
$\left\langle \bar{u}u \right\rangle$ and $\left\langle \bar{d}d \right\rangle$ as well as for the strange quark condensate
$\left\langle \bar{s}s \right\rangle$. We observe that for given value 
of isospin asymmetry parameter, $\eta$,
and strangeness fraction, $f_{s}$, the
magnitude of the values of quark condensate decreases w.r.t. vacuum value.
For example, in symmetric nuclear matter ($\eta$ = 0 and $f_{s}$ = 0),
at nuclear saturation density, $\rho_{B}$ = $\rho_{0}$,
the values of $\left\langle \bar{u}u \right\rangle$, $\left\langle \bar{d}d \right\rangle$ and 
$\left\langle \bar{s}s \right\rangle$
are observed to be $0.629$ $\left\langle \bar{u}u\right\rangle _{0}$, $0.629$ $\left\langle \bar{d}d\right\rangle _{0}$
and $0.895$ $\left\langle \bar{s}s\right\rangle _{0}$ respectively. 
Note that $\left\langle \bar{u}u\right\rangle _{0}$,
 $\left\langle \bar{d}d\right\rangle _{0}$ and
  $\left\langle \bar{s}s\right\rangle _{0}$
denotes the vacuum values of quark condensates and
in chiral SU(3) model these are $-1.401 \times 10^{-2}$ GeV$^{3}$
 and $-1.401 \times 10^{-2}$ GeV$^{3}$ and $-4.671 \times 10^{-2}$ GeV$^{3}$ respectively. 
 As we can see from equations (\ref{qu}), (\ref{qd})
and (\ref{qs}), the values of quark condensates
are proportional to the scalar fields $\sigma$ and
$\zeta$. As discussed above the magnitude of these scalar fields undergo drop
as a function of density of baryonic matter
and this further cause a decrease in the magnitude of quark condensates.
As we move to the asymmetric nuclear matter, say $\eta = 0.5$ and $f_s = 0$,
the values of $\left\langle \bar{u}u\right\rangle $,
 $\left\langle \bar{d}d\right\rangle$ and $\left\langle \bar{s}s\right\rangle$,
at nuclear saturation density, $\rho_{0}$,
are observed to be $0.669$ $\left\langle \bar{u}u\right\rangle _{0}$, $0.607$ $\left\langle \bar{d}d\right\rangle _{0}$
and $0.897$  $\left\langle \bar{s}s\right\rangle _{0}$ respectively. 
Note that as we move to the asymmetric medium the values of
condensate $\left\langle \bar{u}u \right\rangle$
 increases whereas that of $\left\langle \bar{d}d\right\rangle$
decreases due opposite contribution of scalar-isovector
mesons $\delta$.

In isospin symmetric medium ($\eta$ = 0), below baryon density 
$\rho_B$ = $3.3 \rho_0$ as we move from 
non-strange medium, i.e. $f_s$ = 0, to strange medium with 
$f_s$ = 0.5, the magnitude of light quark condensates $\left\langle \bar{u}u\right\rangle $
increases. 
At baryon density $\rho_0$ the values of quark condensate $\left\langle \bar{u}u\right\rangle $
are observed to be 
$0.63$ $\left\langle \bar{u}u\right\rangle _{0}$, $0.643$ $\left\langle \bar{u}u\right\rangle _{0}$
and $0.652$  $\left\langle \bar{u}u\right\rangle _{0}$
 at strangeness fractions $f_s$ = 0, 0.3 and
0.5 respectively. 
Above  baryon density 
$\rho_B$ = 3.3 $\rho_0$ the magnitude of the 
values of light quark condensates is more in 
non strange medium ($f_s$ = 0) as compared 
to strange medium with  finite $f_s$.
At density 4$\rho_0$ these values of quark condensates
changes to $0.3$ $\left\langle \bar{u}u\right\rangle _{0}$, 
$0.291$ $\left\langle \bar{u}u\right\rangle _{0}$
 and $0.292$ $\left\langle \bar{u}u\right\rangle _{0}$ respectively.
 In isospin asymmetric matter with $\eta$ = 0.5 and
 baryon density, $\rho_B$ =  $\rho_0$ the values of   
 light quark condensates $\left\langle \bar{u}u\right\rangle $ are observed to be 
 $0.669$ $\left\langle \bar{u}u\right\rangle _{0}$,
  $0.691 \left\langle  \bar{u}u\right\rangle _{0}$ and 
 $0.708$ $\left\langle \bar{u}u\right\rangle _{0}$ at strangeness fractions 
 $f_s$ = 0, 0.3 and 0.5 respectively. At baryon density
 $4\rho_0$ these values of condensates changes to 
 $0.38$ $\left\langle \bar{u}u\right\rangle _{0}$, 
 $0.379$ $\left\langle \bar{u}u\right\rangle _{0}$
  and $0.392$ $\left\langle \bar{u}u\right\rangle _{0}$ respectively. 
  The values of light quark condensates $\left\langle \bar{d}d\right\rangle $,
  in asymmetric matter with $\eta = 0.5$
  and baryon density $\rho_B = \rho_0$ (4$\rho_0$),
  are observed to be $0.607$ $\left\langle \bar{d}d\right\rangle _{0}$ ($0.285$
  $\left\langle \bar{d}d\right\rangle _{0}$),
   $0.609 \left\langle \bar{d}d \right\rangle _{0}$ 
   ($0.239$ $\left\langle \bar{d}d\right\rangle _{0}$)
    and $0.615$ $\left\langle \bar{d}d\right\rangle _{0}$
    ($0.225$ $\left\langle \bar{d}d\right\rangle _{0}$)
  at strangeness fractions $f_s$ = 0, 0.3 and 0.5
  respectively.
  From figure (\ref{lu}) we observe that as a function of
  density of baryonic matter we always observe a decrease 
  in the values of quark condensates.
  This support the expectation of chiral symmetry 
  restoration at high baryonic density.
  However, in linear Walecka model \cite{quark4,quark5} and also in Dirac-Brueckner
  approach \cite{quark3}  , the values of quark condensates are
  observed to increase at higher values of baryonic
  density and causes hindrance to 
  chiral symmetry restoration. The possible reason for this may be 
  that the chiral invariance property was not considered in
  these calculations \cite{quark3}.
   As the strange quark condensates, $\left\langle \bar{s}s\right\rangle $,
  is proportional to the
  strange scalar-isoscalar field, $\zeta$, 
  therefore the behavior of this field as
  a function of various parameters
  of the medium is also reflected in the
  values of strange quark condensate $\left\langle \bar{s}s\right\rangle $.
  For fixed baryon density, $\rho_B$ and isospin asymmetry
  parameter, $\eta$, as we move from 
  non-strange to strange hadronic medium
  the values of strange condensate  
  decreases. For example, 
  at nuclear saturation density, $\rho_0$
  and isospin asymmetric parameter $\eta = 0$ (0.5),
  the values of $\left\langle \bar{s}s\right\rangle $
  are observed to be $0.895$ $\left\langle \bar{s}s\right\rangle _{0}$ (0.897 $\left\langle \bar{s}s\right\rangle _{0}$), 
  $0.881$ $\left\langle \bar{s}s\right\rangle _{0}$ (0.877 $\left\langle \bar{s}s\right\rangle _{0}$)and
   $0.871$ $\left\langle \bar{s}s\right\rangle _{0}$ (0.863 $\left\langle \bar{s}s\right\rangle _{0}$)
 at $f_s$ = 0, 0.3 and 0.5 respectively.
 At baryonic density $4\rho_0$ and 
 isospin asymmetric parameter $\eta = 0$ (0.5),
  the values of $\left\langle \bar{s}s\right\rangle $
  are observed to be $0.810$ $\left\langle \bar{s}s\right\rangle _{0}$ (0.818 $\left\langle \bar{s}s\right\rangle _{0}$), 
  $0.738$ $\left\langle \bar{s}s\right\rangle _{0}$ (0.717 $\left\langle \bar{s}s\right\rangle _{0}$)and
   $0.684$ $\left\langle \bar{s}s\right\rangle _{0}$ (0.656 $\left\langle \bar{s}s\right\rangle _{0}$)
 at $f_s$ = 0, 0.3 and 0.5 respectively.

\begin{figure}
\includegraphics[width=16cm,height=10cm]{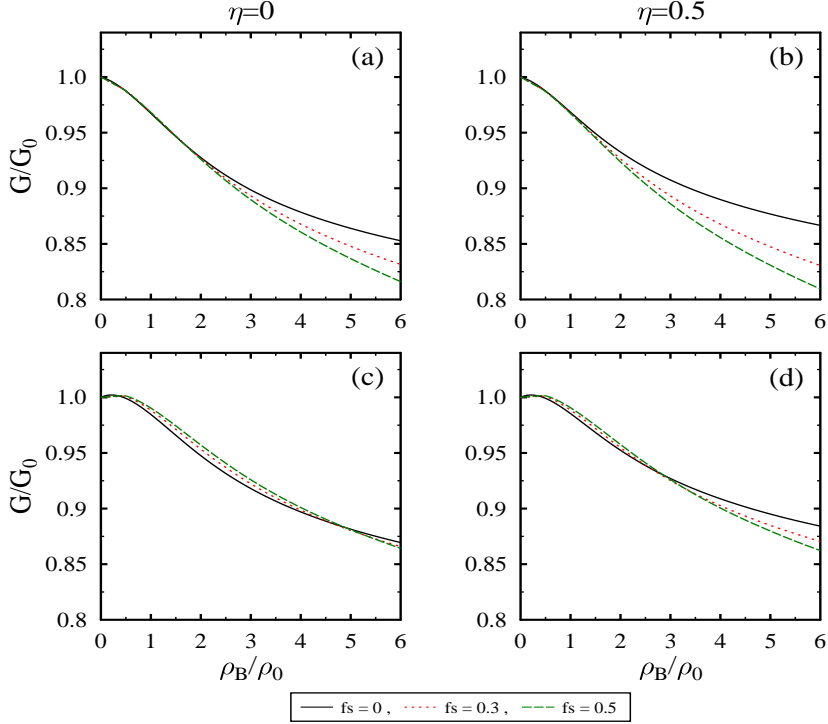}  
\caption{(Color online)
In above figure we plot the ratio of in-medium
scalar gluon condensate to vacuum value of gluon condensate
 as a function of 
 baryonic density, $ \rho_B$ (in units of nuclear saturation density, $\rho_0$). 
 We compare the results at isospin asymmetric parameters $\eta$ = 0 and 0.5. 
 For each value of isospin asymmetry parameter, $\eta$, the results are
 shown for 
 strangeness fractions $f_s$ =0, 0.3 and 0.5. 
 In y-axis $G$ = $\frac{\alpha_s}{\pi}G_{\mu\nu}^{a} G^{\mu\nu a}$. }
\label{gl2}
\end{figure}
  In figure (\ref{gl2}) we show the dependence of Gluon condensates 
  on the density and isospin asymmetry of strange hadronic medium.
  We plot the variation of in-medium gluon condensate
  to vacuum condensate ratio as a function of baryonic
  density of strange hadronic medium.
  In subfigures (a) and (b) we consider the
  contribution of finite quark mass term in the
  calculation of gluon condensates whereas subfigures (c) and (d)
  are
  without the effect of quark mass term (see equation (\ref{tensor4})).
  We observe that as a function of baryonic density of the
  hadronic medium the values of gluon condensates decreases.
 Considering the effect of finite quark mass term,
 at baryonic density, $\rho_B$ = $\rho_0$, and isospin
 asymmetric parameter, $\eta$ = 0 (0.5),
 the values of gluon condensates
 are observed to be $0.985$ $G_{vac}$ ($0.986$ $G_{vac}$), $0.988$ $G_{vac}$ ($0.989$ $G_{vac}$)
 and $0.990$ $G_{vac}$ ($0.991$ $G_{vac}$)
  at strangeness fractions, $f_s$ = 0, 0.3 and
 0.5 respectively.
 However if we do not take into account the
 finite quark mass term
 then 
 for baryonic density, $\rho_B$ = $\rho_0$, and isospin
 asymmetric parameter, $\eta$ = 0 (0.5),
 the values of gluon condensates
 are observed to be 
$0.967$ $G_{vac}$ ($0.968$ $G_{vac}$), $0.968$ $G_{vac}$ ($0.968$ $G_{vac}$)
 and $0.968$ $G_{vac}$ ($0.967$ $G_{vac}$) 
  at strangeness fractions, $f_s$ = 0, 0.3 and
 0.5 respectively.
 As one can see from figures (c) and (d), the effect of
 strangeness fractions are more pronounced at
 higher baryon densities.
 For example, at $\rho_B$ = $4\rho_0$,
 and isospin asymmetry parameter, $\eta = 0$ (0.5),
 the values of gluon condensates are observed
 to be 
$0.879$ $G_{vac}$ ($0.890$ $G_{vac}$), $0.868$ $G_{vac}$ ($0.868$ $G_{vac}$)
 and $0.861$ $G_{vac}$ ($0.856$ $G_{vac}$) 
  at strangeness fractions $f_s$ = 0, 0.3
 and 0.5 respectively. 
The above calculations show that the
gluon condensates have small density dependence as compared to quark condensates.
This observation is consistent with earlier model independent
calculations by Cohen \cite{cohen} and also with the QMC model calculation
in ref. \cite{quark6}.

\begin{figure}
\includegraphics[width=16cm,height=16cm]{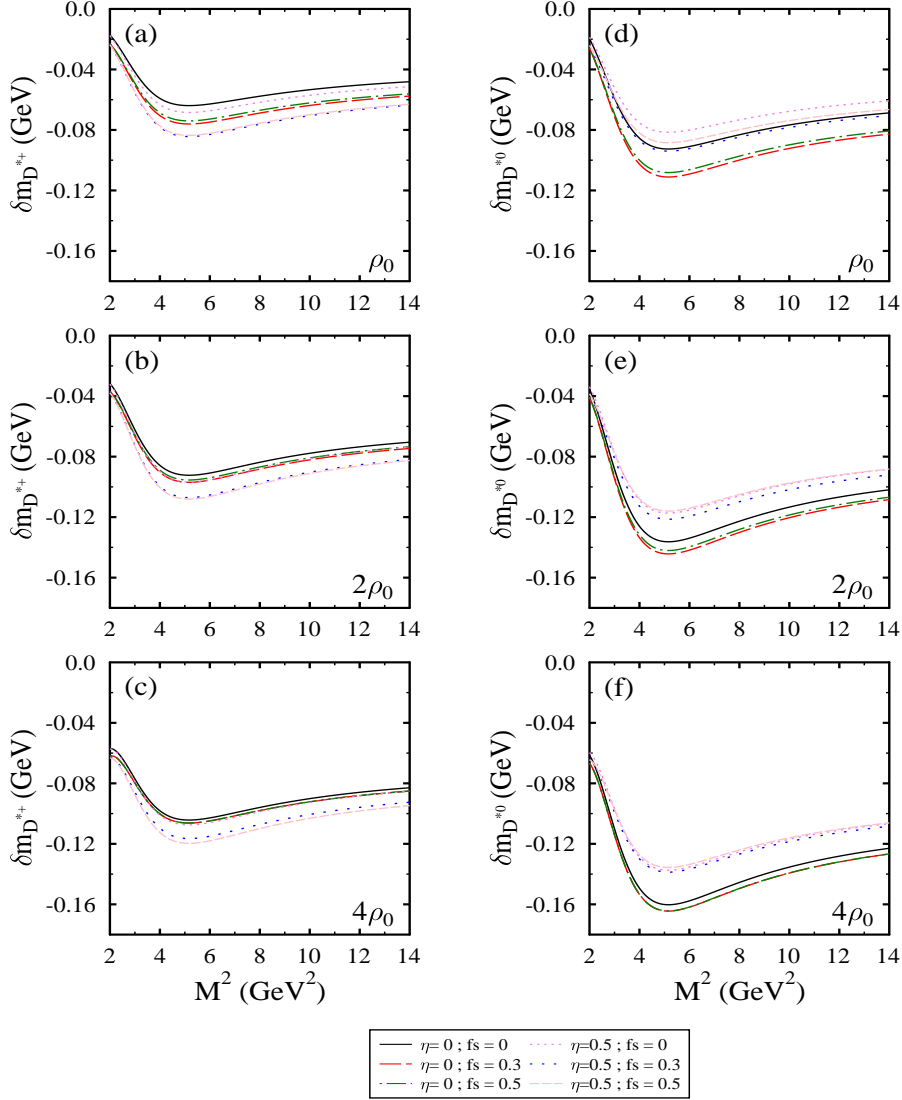}  
\caption{(Color online)
Figure shows the variation of mass shift of
 vector mesons $D^{*+}$ and $D^{*0}$ as a function 
 of squared Borel mass parameter, $M^2$. 
We compare the results at isospin asymmetric parameters $\eta$ = 0 and 0.5. 
 For each value of isospin asymmetry parameter, $\eta$, the results are
 shown for 
 strangeness fractions $f_s$ =0, 0.3 and 0.5.}
\label{dstmass}
\end{figure}

\begin{figure}
\includegraphics[width=16cm,height=16cm]{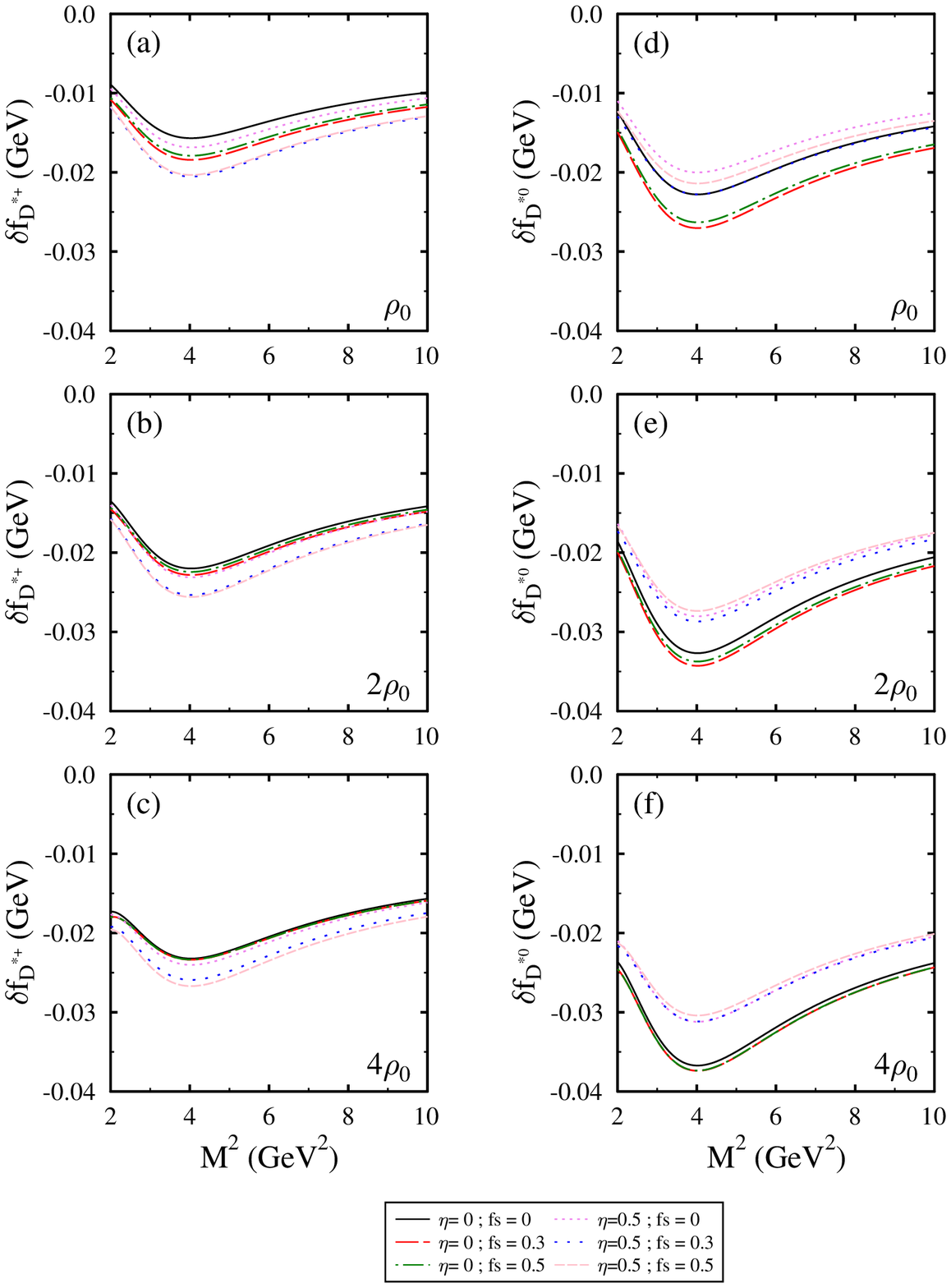}  
\caption{(Color online)
Figure shows the variation of  shift in decay constant of
 vector mesons $D^{*+}$ and $D^{*0}$ as a function 
 of squared Borel mass parameter, $M^2$. 
We compare the results at isospin asymmetric parameters $\eta$ = 0 and 0.5. 
 For each value of isospin asymmetry parameter, $\eta$, the results are
 shown for 
 strangeness fractions $f_s$ =0, 0.3 and 0.5.}
\label{dstdecay}
\end{figure}

 In figures (\ref{dstmass}) and (\ref{dstdecay}) we show the 
 variation of mass shift and shift in decay constant respectively of
  $D^{\star +}$ and $D^{\star 0}$ vector mesons 
  as a function of squared Borel mass parameter, $M^2$.
  In each subplot we  compare the results 
  at $\eta$ = 0 ($f_s$ = 0, 0.3 and 0.5) with $\eta$ = 0.5
  ($f_s$ = 0, 0.3 and 0.5). We present the results at baryon
  densities, $\rho_B$ = $\rho_0$, $2\rho_0$ and $4\rho_0$.

     In symmetric nuclear medium, at nuclear saturation density $\rho_B$ = $\rho_0$,
      the values of mass shifts for $D^{*+}$($D^{*0}$) 
      vector mesons are observed to be -63.8(-92.59) , -76(-111) and -74(-108) MeV for 
       strangeness fractions $f_s$ = 0, 0.3 and 0.5 respectively. 
       The difference in the masses of  $D^{*+}$ and $D^{*0}$
       mesons in the symmetric nuclear medium is due to
       the different masses of $u$ and $d$ quarks
       considered in the present investigation.
       For asymmetric medium($\eta$ = 0.5) and 
        baryonic density, $\rho_B$=$\rho_0$,
         the above values of mass shift changes to -68.4(-81), -84.4(-93.9) 
         and -83.6(-88.5) MeV for 
         strangeness fractions, $f_s$ =0, 0.3 and 0.5 
         respectively. 
From the above quoted
values on mass shift 
we conclude that 
for given value of baryonic density and 
strangeness fraction, 
$D^{*+}$ ($D^{*0}$) mesons undergo large (less) mass drop in
asymmetric medium as compared to symmetric medium.
Note that the $D^{*+}$ meson contain the
light $d$ quark and the $D^{*0}$
meson has light $u$ quark.
As discussed earlier the behavior of $\left\langle\bar{d}d \right\rangle$
and $ \left\langle \bar{u}u\right\rangle$ condensates is
opposite as a function of asymmetry of
the medium and this causes
the observed behavior of  
 $D^{*+}$
and $D^{*0}$ mesons as a function of asymmetry of the
medium.
As compared to nuclear medium ($f_s = 0$)
 the drop in the masses of $D^{*+}$ and $D^{*0}$
mesons is more in strange medium (finite $f_{s}$). 
In symmetric nuclear medium,
at baryon density $\rho_B$ = 2$\rho_0$ $(4\rho_{0})$,
 the values of mass shift for $D^{*+}$ mesons 
 are found to be -92.3(-104.2) , -96.9(-106.25) 
 and -95.4(-106.19) MeV for strangeness fractions
 $f_s$ = 0 ,0.3 and 0.5 respectively. 
In asymmetric matter with $\eta$=0.5 the above values are 
shifted to  -96.7(-107.5), -107(-116.5) and -108(-119.6)  
for strangeness fraction $f_s$ = 0,0.3 and 0.5 respectively. 
At baryon density $\rho_B$ = 2$\rho_0$ $(4\rho_{0})$, 
and $\eta$ = 0.5,
the values of mass shift for $D^{*0}$ mesons
are found to be 
-117(-137), -121(-138.7) and -116 (-135.6) MeV
 for strangeness fraction $f_s$ = 0,0.3 and 0.5 respectively. 
 The drop in the  masses of $D^{*+}$ and $D^{*0}$ mesons
 increases with increase in the baryonic density of the medium.
 
Now we come to the behavior of decay constants of 
$D^{*+}$ and $D^{*0}$ mesons in strange hadronic medium.
 In symmetric nuclear matter ($\eta$ = 0) for baryon density, $\rho_B$ = $\rho_0$,  the  values of shift 
  in decay constants of  $D^{*+}$ ( $D^{*0}$) mesons are observed to be
   -20(-22), -28.2(-28.9) and -28.1(-28.8) MeV for the 
   values of strangeness fractions $f_s$ = 0, 0.3, 0.5 respectively. 
   For asymmetry parameter, $\eta$=0.5, the above values
   of shift in decay constants changes
    to -20.6(-20.7), -28.4(-28.3) and -28.4(-28.1) MeV. 
For density, 2$\rho_0$, and $\eta$ = 0  values of decay shifts are
 -39.7(-40.5), -47.2(-47.9) and -47.1(-47.8) MeV for
  the values of strangeness fractions $f_s$ = 0, 0.3 and 0.5 respectively.
   For the same value of density but $\eta$ = 0.5 above values 
   modified to -39.9(-39.8), -47.5(-47.1) and -47.6(-46.9) MeV respectively. 
Above calculations show that the  presence of
hyperons along with nucleons in the medium   causes more 
decrease in the values of decay constants of
$D^{\star}$ mesons.
 
\begin{figure}
\includegraphics[width=16cm,height=16cm]{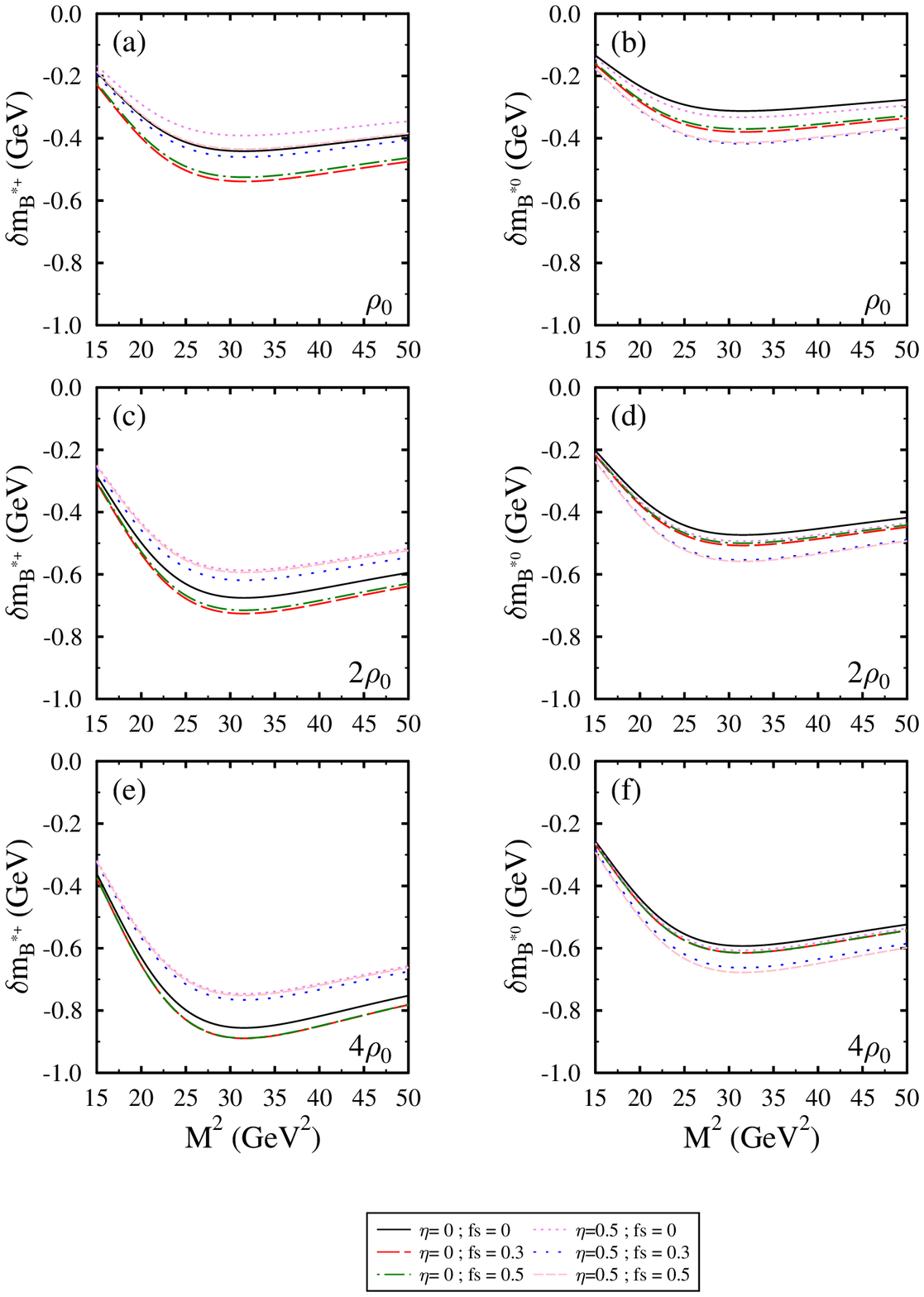}  
\caption{(Color online)
Figure shows the variation of mass shift of
 vector mesons $B^{*+}$ and $B^{*0}$ as a function 
 of squared Borel mass parameter, $M^2$. 
We compare the results at isospin asymmetric parameters $\eta$ = 0 and 0.5. 
 For each value of isospin asymmetry parameter, $\eta$, the results are
 shown for 
 strangeness fractions $f_s$ =0, 0.3 and 0.5.}
\label{Bstmass}
\end{figure}
\begin{figure}
\includegraphics[width=16cm,height=16cm]{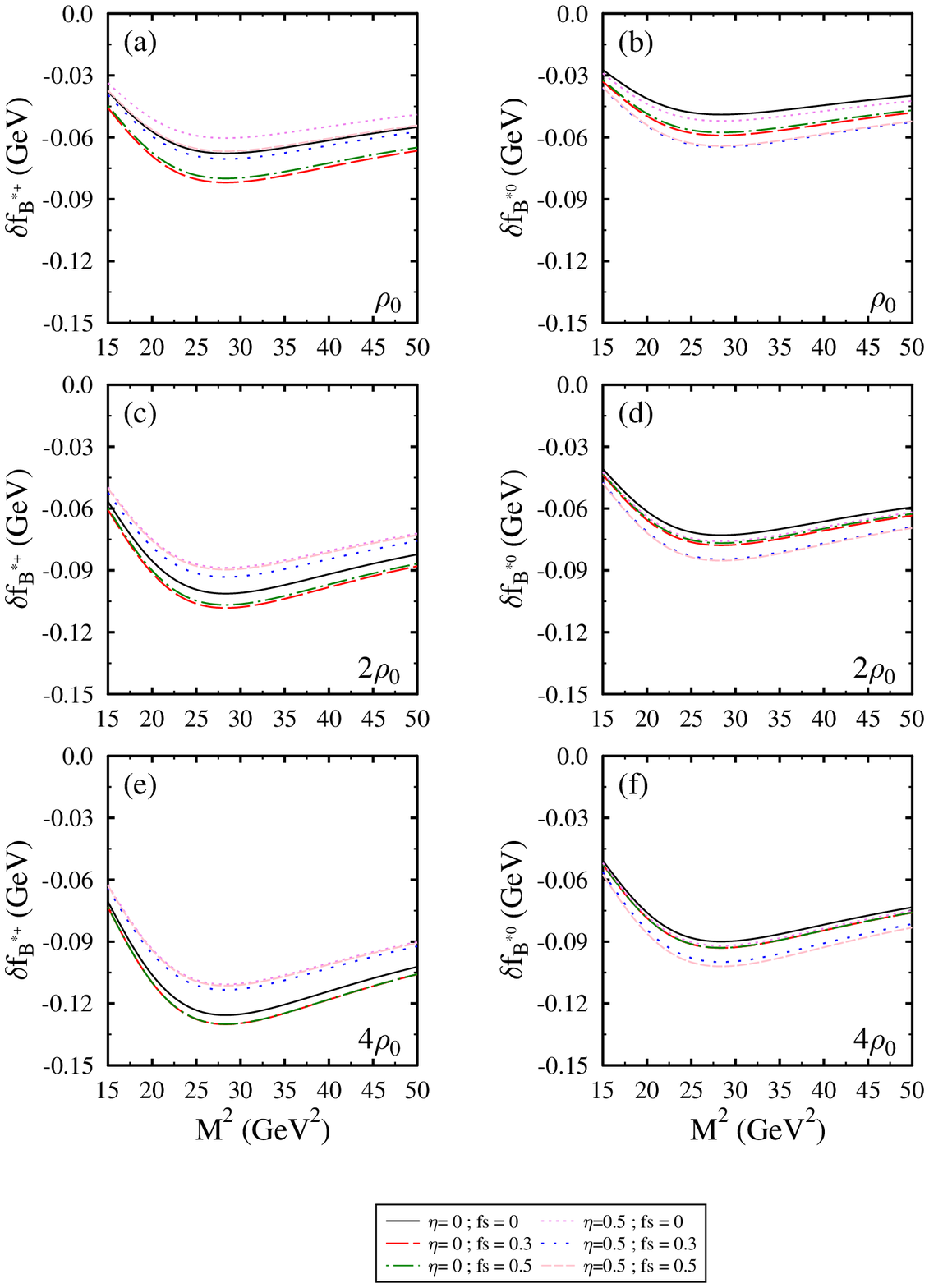}  
\caption{(Color online)
Figure shows the variation of  shift in decay constant of
 vector mesons $B^{*+}$ and $B^{*0}$ as a function 
 of squared Borel mass parameter, $M^2$. 
We compare the results at isospin asymmetric parameters $\eta$ = 0 and 0.5. 
 For each value of isospin asymmetry parameter, $\eta$, the results are
 shown for 
 strangeness fractions $f_s$ =0, 0.3 and 0.5.}
\label{Bstdecay}
\end{figure}

Figures (\ref{Bstmass}) and (\ref{Bstdecay})
show the behaviour of mass shift and decay shift of 
 $B^{*+}$ and $ B^{*0}$ mesons  
 as a function of squared Borel mass parameter
 for different conditions  of the medium.
In table (\ref{Bstmasstable}) and (\ref{Bstdecaytable})
 we tabulate the values of 
  shift in masses and decay constants
  respectively  of $B^{*+}$ and $ B^{*0} $  mesons in nuclear
  and strange hadronic medium. 
  We  observe that for the constant value of 
  strangeness fraction, $f_s$, and isospin 
  asymmetric parameter, $\eta$, the values of mass drop increases as a function 
   of baryonic density. 
If we keep $\eta$ and  $\rho_B$
 constant 
 then the drop in the masses 
  of $B^{*+}$ and $B^{*0}$ mesons is observed to be more 
  at higher  strangeness fractions of the medium. 
  This behavior of vector $B^{*+}$ and $B^{*0}$ mesons 
  as function of density and strangeness fraction of the medium is
  consistent with that of  vector $D^{*+}$ and $D^{*0}$ mesons.
When we keep baryonic density and 
strangeness fraction, $f_s$ constant then 
as compared to symmetric matter the drop in the mass 
of $B^{*+}$ meson is less and that of $B^{*0}$ meson 
is more in asymmetric matter.    
\begin{table}
\begin{tabular}{|p{1cm}|p{1cm}|p{1cm}|p{1.5cm}|p{1.5cm}|p{1.5cm}|p{1.5cm}|}
\hline
 &$\rho_B$&$fs$&\multicolumn{2}{c|}{$\eta$=0}&\multicolumn{2}{c|}{$\eta$=0.5}\\
\cline{4-7}
& & &$B^{*0}$&$B^{*+}$&$B^{*0}$&$B^{*+}$\\
\hline
\multirow{9}{*}{$\delta m_{B^*}$} &   & 0 & -312  & -441 & -332 & -391 \\ 
& $\rho_0$ & 0.3 & -379 & -539 & -417 & -460 \\ 
& &0.5&-370&-525&-413&-435\\ \cline{2-7}

&   & 0 & -473  & -675 & -493 & -587  \\ 
& 2$\rho_0$ & 0.3 & -507 & -726 & -553 & -618 \\  
& &0.5&-499&-715&-558&-594\\ \cline{2-7}

&   & 0 & -592  & -855 & -607 & -746 \\ 
& 4$\rho_0$ & 0.3 & -615 & -889 & -663 & -766 \\  
& &0.5&-614&-888&-677&-752\\ \hline
\end{tabular}
\caption{Table shows the effect of baryonic density $\rho_B$ and isospin asymmetric parameter $\eta$ on the shift in masses (in MeV) of $B^{*0}$ and $B^{*+}$ mesons for different values of strageness fraction $f_s$.}
\label{Bstmasstable}
\end{table}
\begin{table}
\begin{tabular}{|p{1cm}|p{1cm}|p{1cm}|p{1.5cm}|p{1.5cm}|p{1.5cm}|p{1.5cm}|}
\hline
 &$\rho_B$&$fs$&\multicolumn{2}{c|}{$\eta$=0}&\multicolumn{2}{c|}{$\eta$=0.5}\\
\cline{4-7}
& & &$B^{*0}$&$B^{*+}$&$B^{*0}$&$B^{*+}$\\
\hline

  &   & 0        & -48.9  & -67.8 & -52.0 & -60.3 \\ 
  
  &$\rho_0$& 0.3 & -59.0 & -81.9 & -64.7 & -70.4\\ 
  & & 0.5        & -57.6 & -79.9 & -64.1 & -66.7\\  \cline{2-7}  
  
\multirow{3}{*}{$\delta f_{B^*}$}  &   & 0 & -72.9  & -101 & -75.8 & -88.7 \\   
  & 2$\rho_0$ & 0.3 & -77.8 & -108 & -84.6 & -93.1\\ 
  & & 0.5 & -76.7 & -106 & -85.3 & -89.6\\  \cline{2-7}
  
  &  & 0 & -89.9  & -125 & -92 & -110 \\   
  & 4$\rho_0$ & 0.3 & -93 & -130 & -99.8 & -113\\ 
  & & 0.5 & -92.9 & -130 & -102 & -111\\ \hline

\end{tabular}
\caption{Table shows the effect of baryonic density $\rho_B$ and isospin asymmetric parameter $\eta$ on the shift in decay constants (in MeV) of $B^{*0}$ and $B^{*+}$  mesons for different values of strangeness fraction $f_s$.}
\label{Bstdecaytable}
\end{table}
As can be observed from table (\ref{Bstdecaytable}), 
for the constant value of strangeness fraction $f_s$ and
 isospin asymmetric parameter, $\eta$, the drop in the values 
  of decay constant increases with the increase of
   baryonic density of hadronic medium.  
Also the drop in the decay constants of $B^{*+}$ and
$B^{*0}$ mesons is observed to be more in strange medium as
compared to nuclear medium. 
On the other hand when we keep baryonic
 density $\rho_B$  and strangeness fraction
  $f_s$ constant then decrease in the drop of decay shift of $B^{*+}$ mesons 
but an increase in drop of decay shift of $B^{*0}$ 
is observed with increasing isospin asymmetric parameter $\eta$.  
\begin{figure}
\includegraphics[width=16cm,height=16cm]{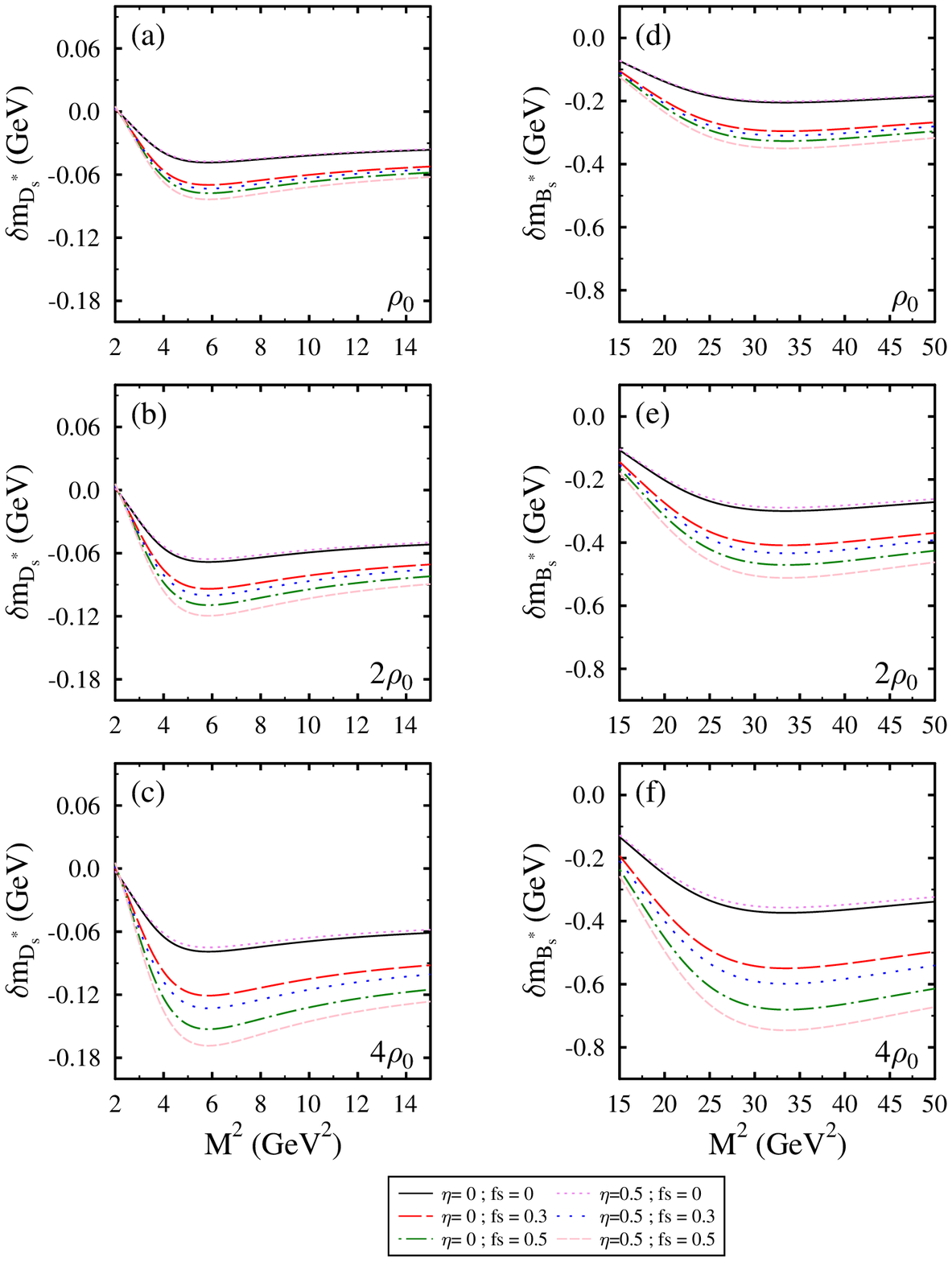}  
\caption{(Color online)
Figure shows the variation of mass shift of
 strange vector mesons $D_{s}^{*}$ and $B_{s}^{*}$ as a function 
 of squared Borel mass parameter, $M^2$. 
We compare the results at isospin asymmetric parameters $\eta$ = 0 and 0.5. 
 For each value of isospin asymmetry parameter, $\eta$, the results are
 shown for 
 strangeness fractions $f_s$ =0, 0.3 and 0.5.}
\label{DsBsmass}
\end{figure}

\begin{figure}
\includegraphics[width=16cm,height=16cm]{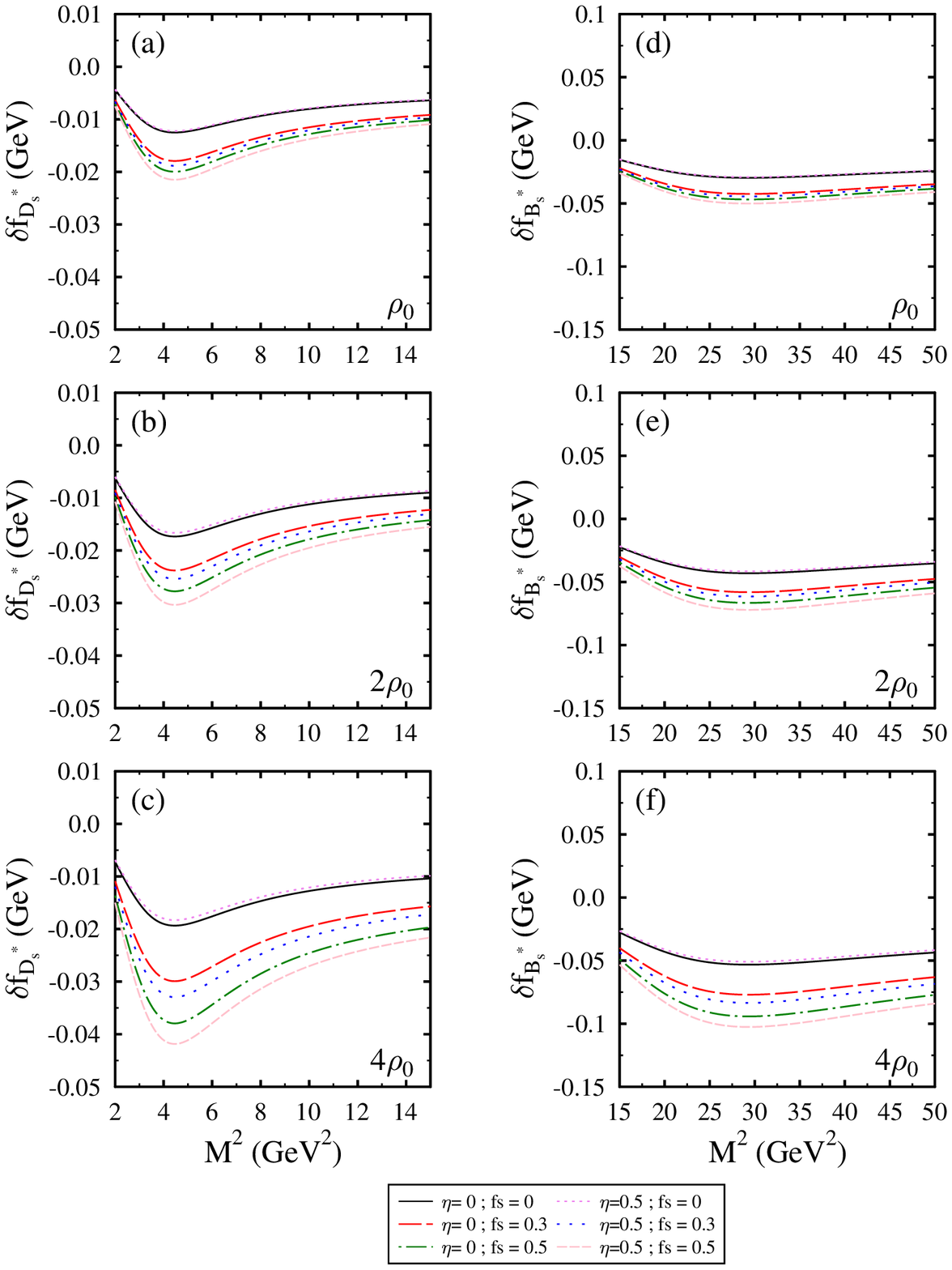}  
\caption{(Color online)
Figure shows the variation of  shift in decay constant of
 strange vector mesons $D_{s}^{*}$ and $B_{s}^{*}$ as a function 
 of squared Borel mass parameter, $M^2$. 
We compare the results at isospin asymmetric parameters $\eta$ = 0 and 0.5. 
 For each value of isospin asymmetry parameter, $\eta$, the results are
 shown for 
 strangeness fractions $f_s$ =0, 0.3 and 0.5.}
\label{DsBsdecay}
\end{figure}

In the present work we also studied the effect
 of strangeness fraction and isospin asymmetric parameter 
of hadronic medium  on the mass shift and decay shift of strange 
charmed and bottom vector mesons $D_s^*$ and $B_s^*$
respectively. In figures (\ref{DsBsmass}) and
(\ref{DsBsdecay}) we present the
variations of mass shift and shift in decay constant respectively of $D_s^*$
 and $B_s^*$ mesons.
 The charmed strange and bottom strange mesons have one strange, $s$, quark
 and one heavy quark.
 The properties of these mesons are calculated in the
 medium through the presence of strange quark
 condensate, $\left\langle \bar{s}s\right\rangle $ in operator product expansion
 side of QCD sum rule equations (\ref{qcdsumdst}) and (\ref{qcdsumd1}).
 In symmetric medium i.e $\eta$ = 0  and for 
 baryon density, $\rho_B$ = $\rho_0$,
  the values of mass shift of $D_s^*$($B_s^*$) 
  mesons are observed to be -48.4(-204),-69.7(-295) and
   -77.6(-326) MeV at strangeness fractions,
    $f_s$ = 0, 0.3 and 0.5 respectively.  
    For density, $\rho_B$ = 2$\rho_0$, the
     above values are found to be -68.3(-299), -93.9(-408)
      and -109(-470) MeV respectively.
   From above discussion we observe that the  strange charmed and bottom vector
   meson undergo mass drop in the hadronic medium.
   Also the values of mass drop is observed to be
   more in strange hadronic medium as compared to 
   non-strange medium. 
In asymmetric parameter, $\eta$ = 0.5 and for 
density, $\rho_B$ = $\rho_0$, the values of mass shift of
 $D_s^*$($B_s^*$) mesons are observed to be -47.4(-200), 
 -73.3(-309) and -83.5(-350) MeV for the strangeness
  fractions $f_s$ = 0, 0.3 and 0.5 respectively. 
   As we move to more dense medium 
   i.e. $\rho_B$ = 2$\rho_0$ then above
    shift in masses are observed to be -65.7(-289),
     -100(-433) and -119(-511) MeV for strangeness
      fraction $f_s$ = 0,0.3 and 0.5 respectively. 
  Thus we notice that  the effect of increasing the
  density or strangeness fraction or isospin asymmetry of the
  medium is to decrease 
    the masses of  $D_s^*$  and $B_s^*$ mesons.

In figure (9) we have shown the effect of strangeness fraction and isospin asymmetric parameter on the decay constant of $D_s^*$($B_s^*$) mesons. In symmetric medium, at nuclear saturation density, $\rho_B$ = $\rho_0$, the values of  shift in decay constants 
of  $D_s^*$($B_s^*$) mesons are 
 found to be -12.5(-29.7), -17.9(-42.4) and -19.9(-46.8) MeV at strangeness fractions $f_s$ = 0, 0.3 and 0.5 respectively.  This indicate that with increase in the strangeness fraction the values of decay constants decrease more from its vacuum value. It is also noticed  that with increase in the baryonic density magnitude of shift in decay constant increases more e.g. at $\rho_B$ = 2$\rho_0$  the
above values altered to -17.3(-43.0), -23.8(-58.0) and -27.7(-66.5) MeV. In nuclear medium when we move from symmetric medium ($\eta$=0) to asymmetric medium($\eta$=0.5) we observed slight decrease in the magnitudes of decay shift of $D_s^*$ ($B_s^*$) mesons. However, in strange hadronic medium ($f_s$=0.5) when we move from symmetric to asymmetric medium, increase in the magnitudes of decay shift is observed. For example, at nuclear saturation density, $\rho_B$ = $\rho_0$ the values of shift in decay constant of $D_s^*$ ($B_s^*$) mesons are found to be  -12.2(-29.1), -18.8(-44.4) and -21.5(-50.1) MeV at strangeness fractions, $f_s$ = 0, 0.3 and 0.5 respectively.

It may be noted that among the various condensates
present in the QCD sum rule equations (\ref{qcdsumdst}) and
(\ref{qcdsumd1}), the scalar quark condensates, $\left\langle \bar{q}q\right\rangle $ have
 largest contribution for the medium modification of $D$ and $B$ meson properties.
 For example, if
 all the condensates are set to zero except 
$<\bar{q}q>$ then shift in mass (decay constant) 
for $D^{*+}$ meson in  symmetric nuclear
medium ($fs$ = 0 and $\eta$ = 0) is observed to
 be -69(-17.8) MeV for $\rho_B$ = $\rho_0$
 and can be compared to the values   -63.2 (-20) MeV
 evaluated in the presence of all condensates.
Also the condensate $<q^\dag \iota D_0 q>$ is not
evaluated within chiral SU(3) model. However its
contribution to the properties of $D$ and $B$ mesons
is very small.
Neglecting $<q^\dag \iota D_0 q>$ only, the values of mass shift and decay shift
in symmetric medium, at $\rho_B$ = $\rho_0$,
are
 observed to be -62 and -15 MeV respectively.

\begin{table}
\begin{tabular}{|p{1cm}|p{1cm}|p{1cm}|p{1.5cm}|p{1.5cm}|p{1.5cm}|p{1.5cm}|}
\hline
 &$\rho_B$&$fs$&\multicolumn{2}{c|}{$\eta$=0}&\multicolumn{2}{c|}{$\eta$=0.5}\\
\cline{4-7}
& & &$D_1^{0}$&$D_1^{+}$&$D_1^{0}$&$D_1^{+}$\\
\hline
\multirow{9}{*}{$\delta m_{D_1}$} &   & 0 &88  &62 & 78 & 69 \\ 
& $\rho_0$ & 0.3 &106  & 74 & 91 & 82 \\ 
& &0.5&103&72&85&81\\ \cline{2-7}

&   & 0 & 130  & 91 & 114 & 95  \\ 
& 2$\rho_0$ & 0.3 & 139 & 97 & 119 & 106 \\  
& &0.5&136&96&114&106\\ \cline{2-7}

&   & 0 & 158  & 110 & 139 & 112 \\ 
& 4$\rho_0$ & 0.3 & 163 & 113 & 142 & 122 \\  
& &0.5&163&113&139&124\\ \hline
\end{tabular}
\caption{Table shows values of mass shift
 (in MeV) of  $D_1^{0}$ and $D_1^{+}$ axial vector mesons
 at different values of baryonic density and for different values of
 isospin asymmetry, $\eta$, and  strangeness fraction, $f_s$,
  of hadronic medium.}
  \label{D1masstable}
\end{table}

\begin{figure}
\includegraphics[width=16cm,height=16cm]{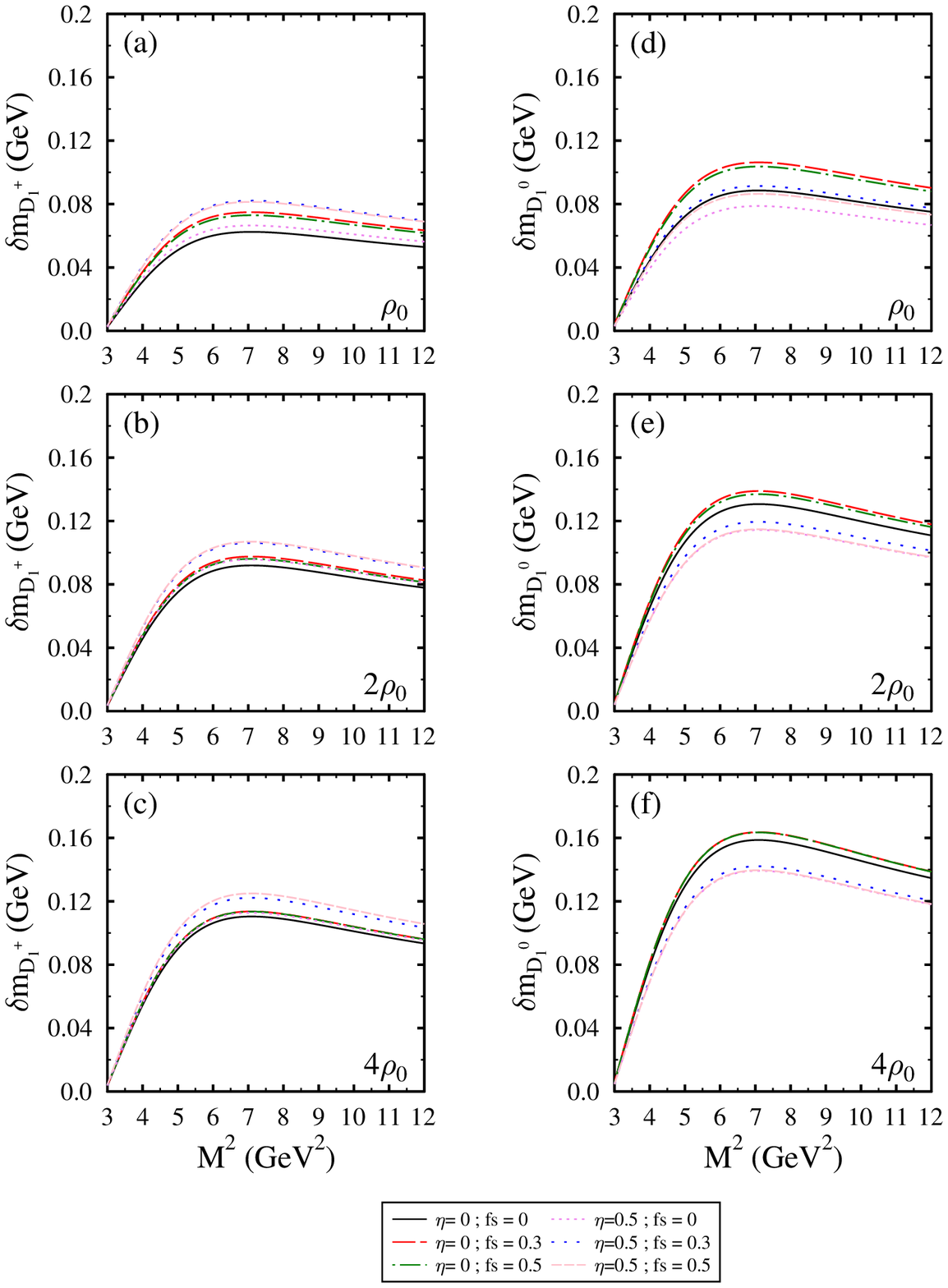}  
\caption{(Color online)
Figure shows the variation of mass shift of
 axial-vector mesons $D_{1}^{+}$ and $D_{1}^{0}$ as a function 
 of squared Borel mass parameter, $M^2$. 
We compare the results at isospin asymmetric parameters $\eta$ = 0 and 0.5. 
 For each value of isospin asymmetry parameter, $\eta$, the results are
 shown for 
 strangeness fractions $f_s$ =0, 0.3 and 0.5.}
\label{D1mass}
\end{figure}

\begin{figure}
\includegraphics[width=16cm,height=16cm]{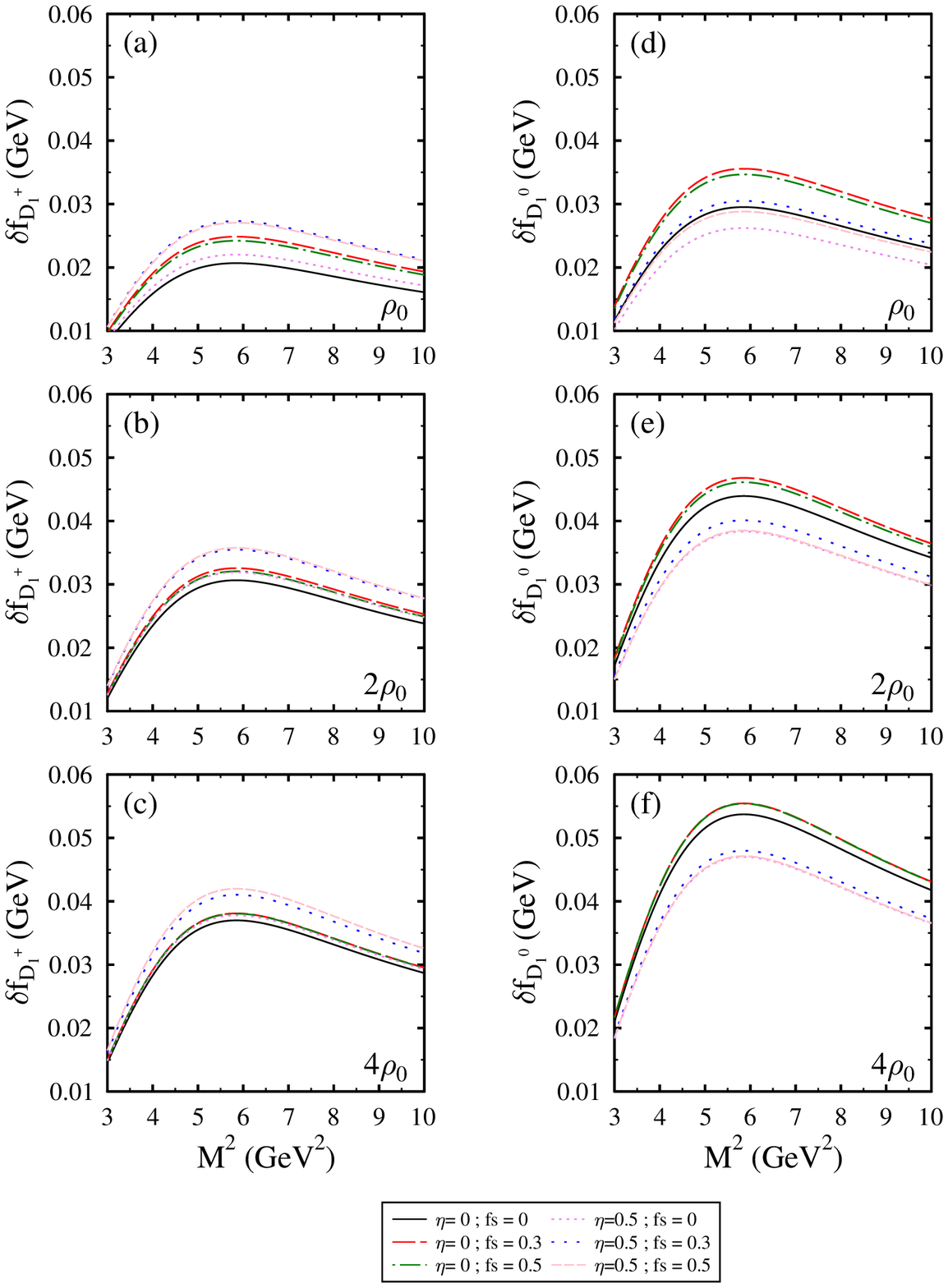}  
\caption{(Color online)
Figure shows the variation of  shift in decay constant of
 axial-vector mesons $D_{1}^{+}$ and $D_{1}^{0}$ as a function 
 of squared Borel mass parameter, $M^2$. 
We compare the results at isospin asymmetric parameters $\eta$ = 0 and 0.5. 
 For each value of isospin asymmetry parameter, $\eta$, the results are
 shown for 
 strangeness fractions $f_s$ =0, 0.3 and 0.5.}
\label{D1decay}
\end{figure}

\begin{figure}
\includegraphics[width=16cm,height=16cm]{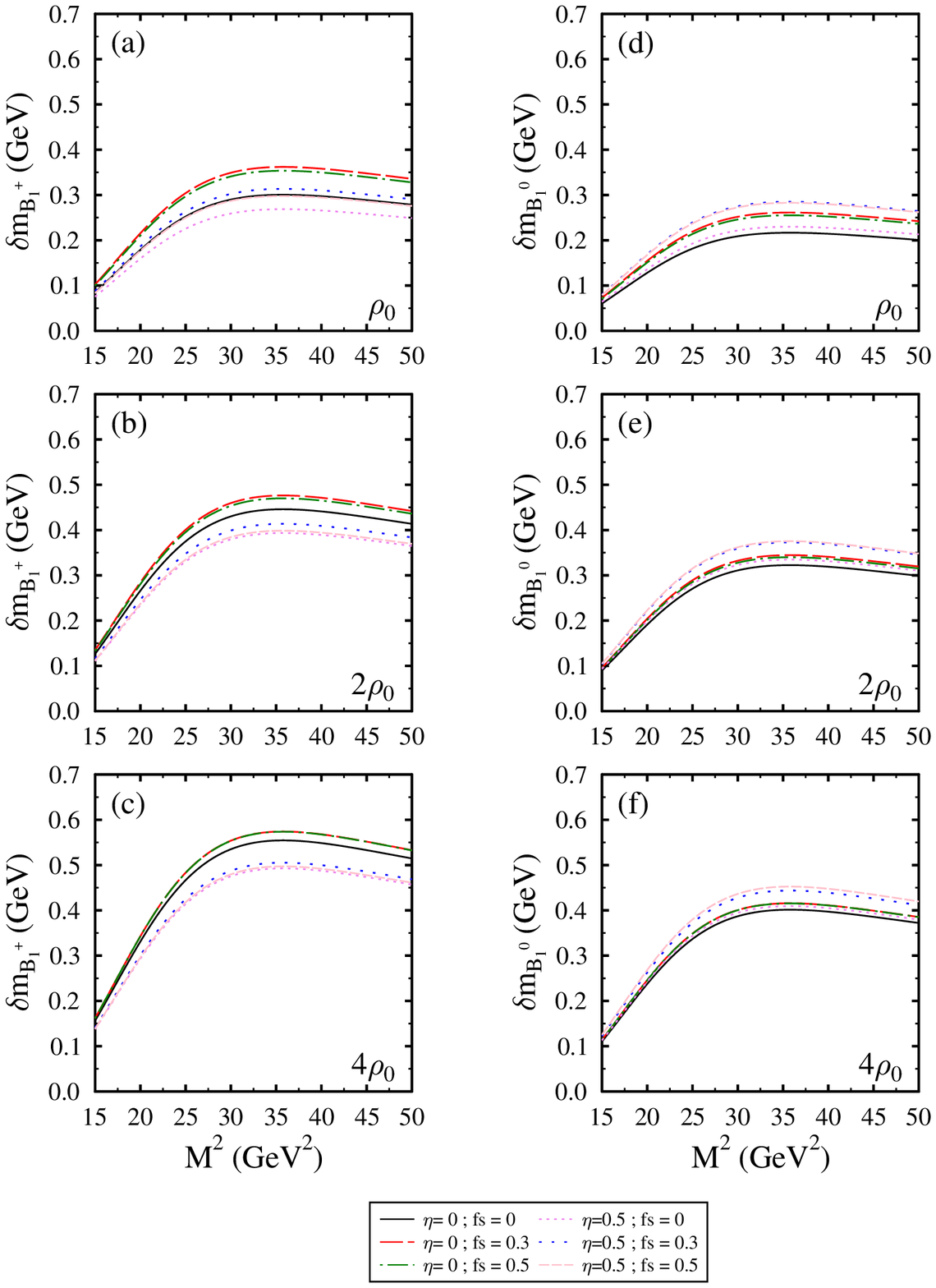}  
\caption{(Color online)
Figure shows the variation of mass shift of
 axial-vector mesons $B_{1}^{+}$ and $B_{1}^{0}$ as a function 
 of squared Borel mass parameter, $M^2$. 
We compare the results at isospin asymmetric parameters $\eta$ = 0 and 0.5. 
 For each value of isospin asymmetry parameter, $\eta$, the results are
 shown for 
 strangeness fractions $f_s$ =0, 0.3 and 0.5.}
\label{B1mass}
\end{figure}

\begin{figure}
\includegraphics[width=16cm,height=16cm]{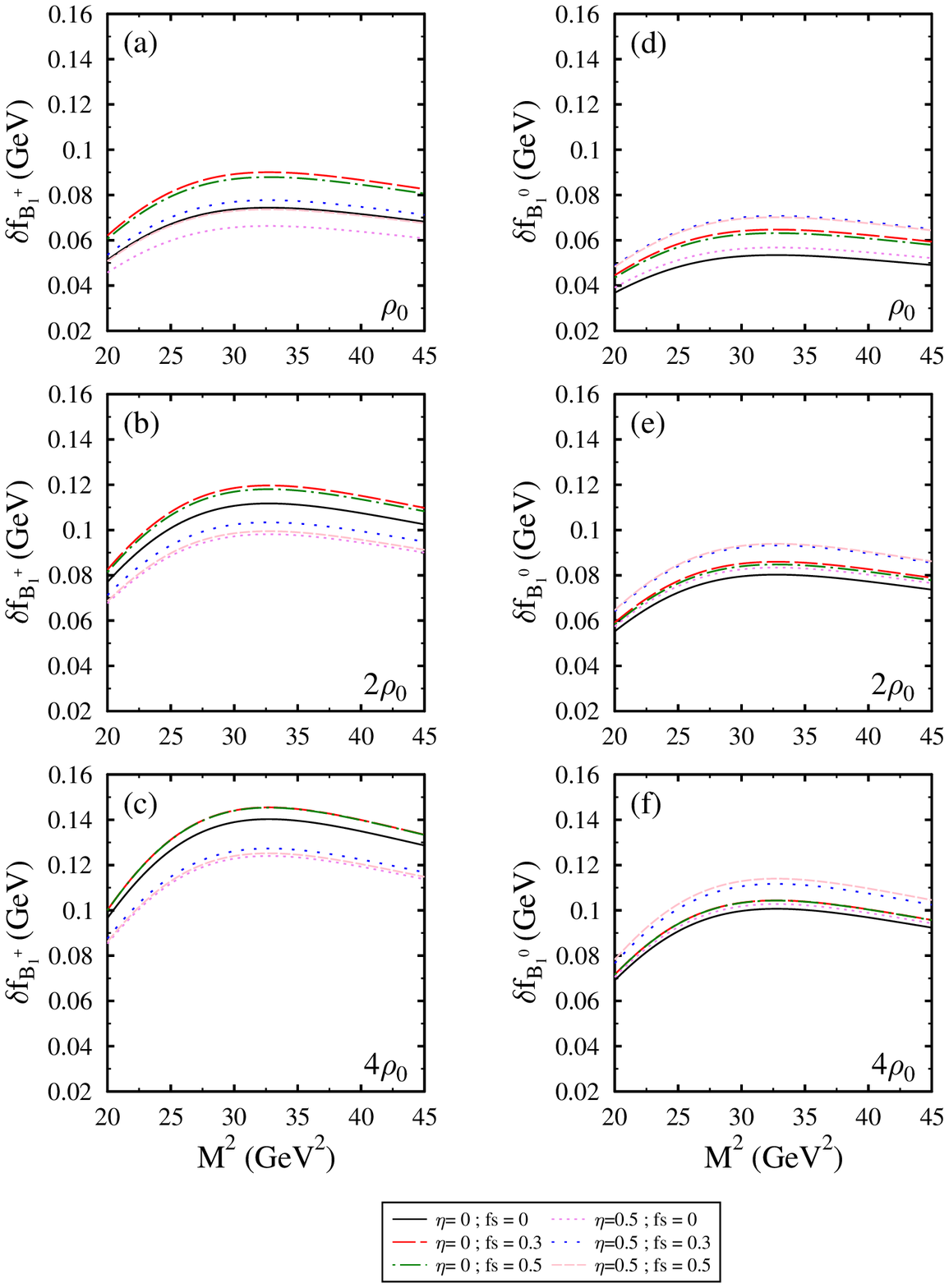}  
\caption{(Color online)
Figure shows the variation of  shift in decay constant of
 axial-vector mesons $B_{1}^{+}$ and $B_{1}^{0}$ as a function 
 of squared Borel mass parameter, $M^2$. 
We compare the results at isospin asymmetric parameters $\eta$ = 0 and 0.5. 
 For each value of isospin asymmetry parameter, $\eta$, the results are
 shown for 
 strangeness fractions $f_s$ =0, 0.3 and 0.5.}
\label{B1decay}
\end{figure}

Figures (\ref{D1mass}) and (\ref{B1mass})  show the modification 
of mass shift  of axial vector $D_1$ ($D_1^{0}$ and $D_1^{+}$)
and $B_1$ ($B_1^{0}$ and $B_1^{+}$) mesons respectively
and figures (\ref{D1decay}) and (\ref{B1decay}) show
the decay shift of these mesons
as a function of squared Borel mass parameter i.e $M^2$
 for different values of baryonic density $\rho_B$ ,
  strangeness fraction $f_s$ and isospin asymmetric parameter i.e $\eta$. 
   In tables (\ref{D1masstable}), (\ref{D1decaytable}),
   (\ref{B1masstable}), (\ref{B1decaytable}) 
   we have given some numerical data
    to discuss the  modification of masses and decay constants of
    axial vector mesons.
    For a constant value of strangeness fraction, $f_s$,
      and isospin asymmetric parameter, $\eta$, as
      a function of baryonic density 
      a positive shift in masses and decay constants
        of axial vector $D_1$ ($D_1^{0}$ and $D_1^{+}$)and $B_1$ 
        ($B_1^{0}$ and $B_1^{+}$) mesons was
        observed.     
     For given density and isospin asymmetry, the
      finite strangeness fraction of the medium also causes an 
      increase in the masses and decay constants 
      of above axial vector mesons.
  For the axial vector $D_1$ and $B_1$ meson
   doublet, as a function of isospin asymmetry of the medium 
       the values of mass shift and decay shift of $D_1^{0}$ ($B_1^{0}$) 
       meson decreases (increases) whereas that
       of  $D_1^{+}$ ($B_1^{+}$)  increases (decreases).
       As discussed earlier in case of vector mesons,
       the reason for opposite behavior of $D$ and $B$ mesons
       as a function of isospin asymmetry of medium is the
       presence of light $u$ quark  in 
       $D_1^{0}$ and $B_1^{+}$ mesons whereas  
       the mesons $D_1^{+}$ and $B_1^{0}$ have light $d$ quark.
              
\begin{table}
\begin{tabular}{|p{1cm}|p{1cm}|p{1cm}|p{1.5cm}|p{1.5cm}|p{1.5cm}|p{1.5cm}|}
\hline
&$\rho_B$&$fs$&\multicolumn{2}{c|}{$\eta$=0}&\multicolumn{2}{c|}{$\eta$=0.5}\\
\cline{4-7}
& & &$D_1^{0}$&$D_1^{+}$&$D_1^{0}$&$D_1^{+}$\\
\hline
\multirow{9}{*}{$\delta f_{D_1}$} &   & 0 &29  &20 & 29 & 22 \\ 
& $\rho_0$ & 0.3 &35  & 24 & 30 & 27 \\ 
& &0.5&34&24&28&27\\ \cline{2-7}

&   & 0 & 43  & 30 & 38 & 31  \\ 
& 2$\rho_0$ & 0.3 & 46 & 32 & 40 & 35 \\  
& &0.5&46&32&38&35\\ \cline{2-7}

&   & 0 & 53  & 37& 47 & 37 \\ 
& 4$\rho_0$ & 0.3 & 55 & 36 & 48 & 40 \\  
& &0.5&55&37&47&42\\ \hline
\end{tabular}
\caption{Table shows the effect of baryonic density $\rho_B$ and isospin asymmetric parameter $\eta$ on the shift in decay constants (in MeV) of $D_1^{0}$ and $D_1^{+}$ mesons for different values of strangeness fraction $f_s$}
\label{D1decaytable}
\end{table}

\begin{table}
\begin{tabular}{|p{1cm}|p{1cm}|p{1cm}|p{1.5cm}|p{1.5cm}|p{1.5cm}|p{1.5cm}|}
\hline
&$\rho_B$&$fs$&\multicolumn{2}{c|}{$\eta$=0}&\multicolumn{2}{c|}{$\eta$=0.5}\\
\cline{4-7}
& & &$B_1^{0}$&$B_1^{+}$&$B_1^{0}$&$B_1^{+}$\\
\hline
\multirow{9}{*}{$\delta m_{B_1}$} &   & 0 &216  &300 & 229 & 268 \\ 
& $\rho_0$ & 0.3 &261  & 362 & 285 & 313 \\ 
& &0.5&255&353&282&297\\ \cline{2-7}

&   & 0 & 322  & 445 & 334 & 393  \\ 
& 2$\rho_0$ & 0.3 & 344 & 476 & 372 & 413 \\  
& &0.5&339&470&375&398\\ \cline{2-7}

&   & 0 & 401  & 554& 409 & 492 \\ 
& 4$\rho_0$ & 0.3 & 415 & 574 & 443 & 505 \\  
& &0.5&415&573&452&497\\ \hline
\end{tabular}
\caption{Table shows the effect of baryonic density $\rho_B$ and isospin asymmetric parameter $\eta$ on the shift in masses(in MeV) of $B_1^{0}$ and $B_1^{+}$ mesons for different values of strangeness fraction $f_s$}
\label{B1masstable}
\end{table}

\begin{table}
\begin{tabular}{|p{1cm}|p{1cm}|p{1cm}|p{1.5cm}|p{1.5cm}|p{1.5cm}|p{1.5cm}|}
\hline
&$\rho_B$&$fs$&\multicolumn{2}{c|}{$\eta$=0}&\multicolumn{2}{c|}{$\eta$=0.5}\\
\cline{4-7}
& & &$B_1^{0}$&$B_1^{+}$&$B_1^{0}$&$B_1^{+}$\\
\hline
\multirow{9}{*}{$\delta f_{B_1}$} &   & 0 &53  &74 & 56 & 66 \\ 
& $\rho_0$ & 0.3 &64  & 90 & 70 & 77 \\ 
& &0.5&63&87&70&73\\ \cline{2-7}

&   & 0 & 80  & 117 & 83 & 98  \\ 
& 2$\rho_0$ & 0.3 & 85 & 119 & 93 & 103 \\  
& &0.5&84&118&93&99\\ \cline{2-7}

&   & 0 & 100  & 140& 102 & 124 \\ 
& 4$\rho_0$ & 0.3 & 104 & 145 & 111 & 127 \\  
& &0.5&104&145&113&125\\ \hline
\end{tabular}
\caption{Table shows the effect of baryonic density $\rho_B$ and isospin asymmetric parameter $\eta$ on the shift in decay constants (in MeV) of $B_1^{0}$ and $B_1^{+}$ mesons for different values of strangeness fraction $f_s$.}
\label{B1decaytable}
\end{table}


\begin{table}
\begin{tabular}{|p{1cm}|p{1cm}|p{1cm}|p{1.5cm}|p{1.5cm}|p{1.5cm}|p{1.5cm}|}
\hline
&$\rho_B$&$fs$&\multicolumn{2}{c|}{$\eta$=0}&\multicolumn{2}{c|}{$\eta$=0.5}\\
\cline{4-7}
& & &$D_{1S}$&$B_{1S}$&$D_{1S}$&$B_{1S}$\\
\hline
\multirow{9}{*}{$\delta m_{D_{1s}}$} &   & 0 &44 &164 & 43 & 161 \\ 
& $\rho_0$ & 0.3 &63  &233 & 65 & 244 \\ 
& &0.5&69&257&74&274\\ \cline{2-7}

&   & 0 & 65  & 238 & 62 & 230  \\ 
& 2$\rho_0$ & 0.3 & 86 & 319 & 91 & 337 \\  
& &0.5&97&363&107&391\\ \cline{2-7}

&   & 0 & 81  & 297& 78 & 285 \\ 
& 4$\rho_0$ & 0.3 & 115 & 422 & 124 & 456 \\  
& &0.5&140&510&151&553\\ \hline
\end{tabular}
\caption{Table shows the effect of baryonic density
 $\rho_B$ and isospin asymmetric parameter $\eta$
  on the shift in masses (in MeV) of $B_{1S}$ and
   $D_{1S}$ mesons for different values of strangeness fraction $f_s$.}
\label{D1smassshifttable}
\end{table}

\begin{figure}
\includegraphics[width=16cm,height=16cm]{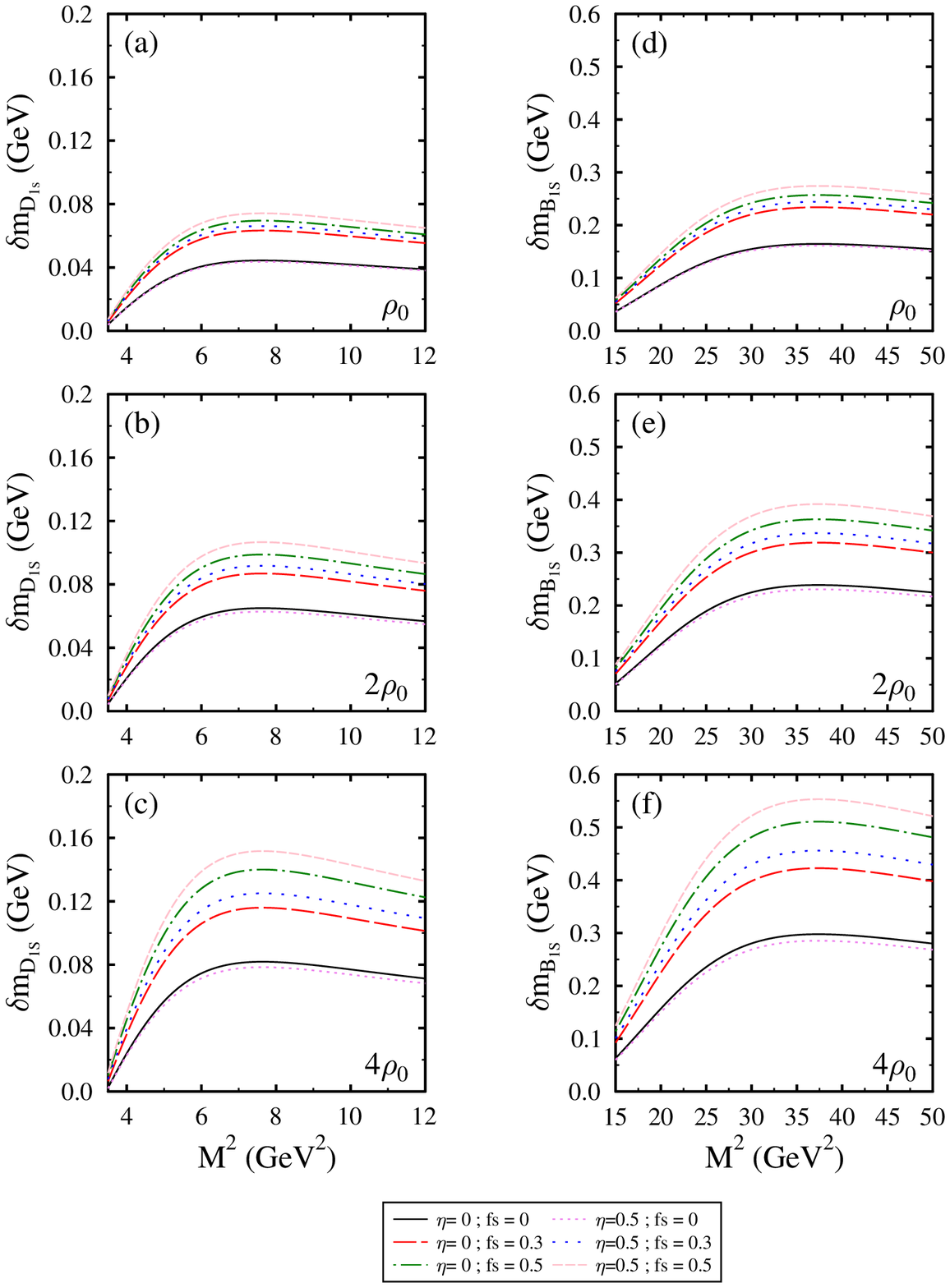}  
\caption{(Color online)
Figure shows the variation of mass shift of
 strange axial-vector mesons $D_{1s}$ and $B_{1s}$ as a function 
 of squared Borel mass parameter, $M^2$. 
We compare the results at isospin asymmetric parameters $\eta$ = 0 and 0.5. 
 For each value of isospin asymmetry parameter, $\eta$, the results are
 shown for 
 strangeness fractions $f_s$ =0, 0.3 and 0.5.}
\label{Ds1Bs1mass}
\end{figure}

\begin{figure}
\includegraphics[width=16cm,height=16cm]{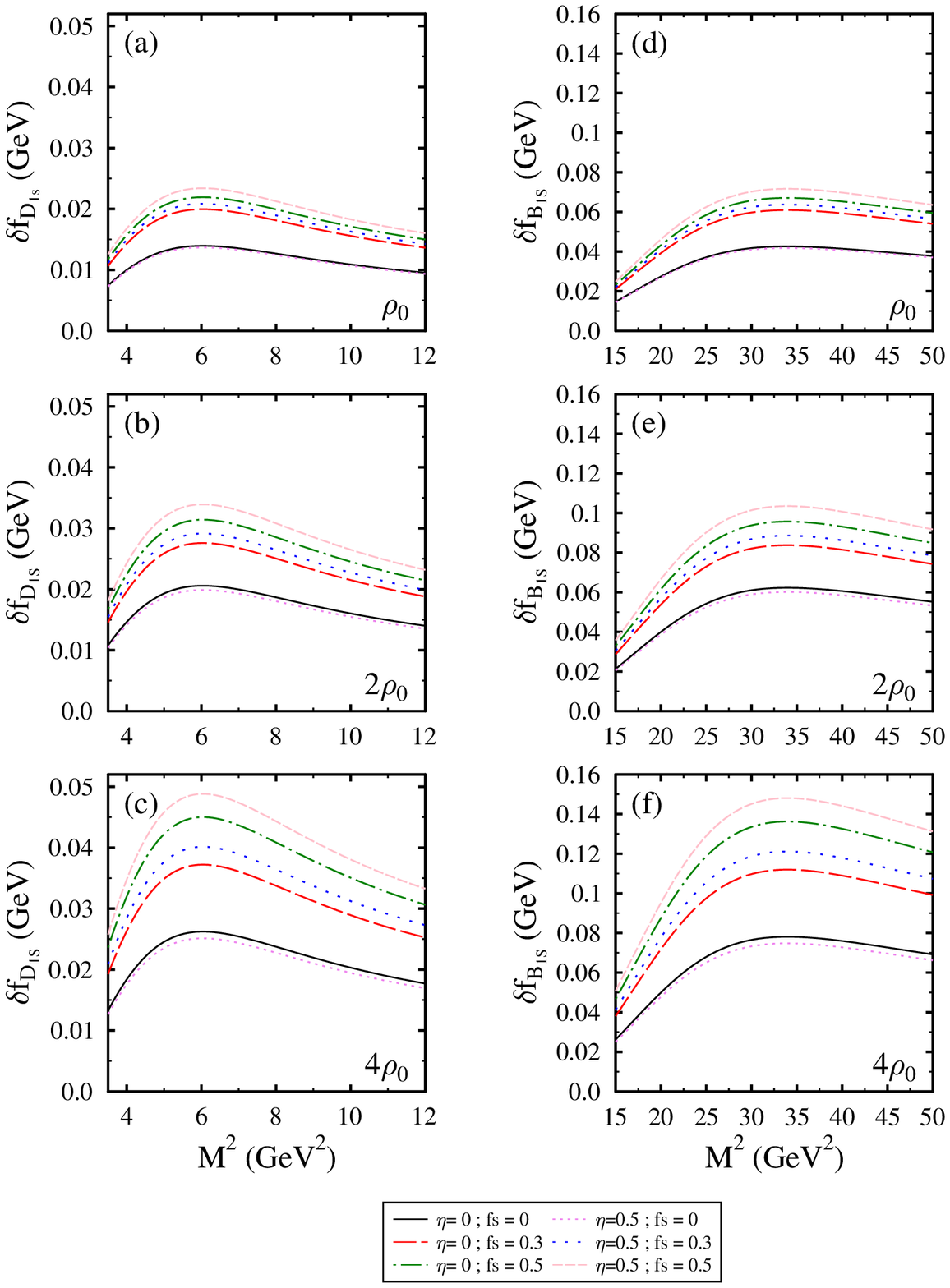}  
\caption{(Color online)
Figure shows the variation of  shift in decay constant of
 strange axial-vector mesons $D_{1s}$ and $B_{1s}$ as a function 
 of squared Borel mass parameter, $M^2$. 
We compare the results at isospin asymmetric parameters $\eta$ = 0 and 0.5. 
 For each value of isospin asymmetry parameter, $\eta$, the results are
 shown for 
 strangeness fractions $f_s$ =0, 0.3 and 0.5.}
\label{Ds1Bs1decay}
\end{figure}

Figure (\ref{Ds1Bs1mass}) and (\ref{Ds1Bs1decay})
 shows the effect of strangeness fraction, isospin
  asymmetric parameter and baryonic density on the
mass shift and decay shift respectively of $D_{1S}$ and $B_{1S}$ mesons.
In tables (\ref{D1smassshifttable}) and (\ref{D1sdecayshifttable})
we tabulate some values of mass shift 
and decay shift for these strange axial vector $D_{1S}$ and $B_{1S}$ mesons.
At  finite baryonic density of the
medium, an increase in strangeness fraction or 
isospin asymmetry of the medium
causes an increase in the positive mass shift and decay
shift of $D_{1S}$ and $B_{1S}$ mesons.

\begin{table}
\begin{tabular}{|p{1cm}|p{1cm}|p{1cm}|p{1.5cm}|p{1.5cm}|p{1.5cm}|p{1.5cm}|}
\hline
&$\rho_B$&$fs$&\multicolumn{2}{c|}{$\eta$=0}&\multicolumn{2}{c|}{$\eta$=0.5}\\
\cline{4-7}
& & &$D_{1S}$&$B_{1S}$&$D_{1S}$&$B_{1S}$\\
\hline
\multirow{9}{*}{$\delta f_{D_{1s}}$} &   & 0 &14  &42 & 13 &41  \\ 
& $\rho_0$ & 0.3 &20  & 60 & 20 & 63 \\ 
& &0.5&22&67&23&71\\ \cline{2-7}

&   & 0 & 20  & 62 & 20 &60  \\ 
& 2$\rho_0$ & 0.3 &  27 &83 & 29 & 88 \\  
& &0.5&31&95&34&103\\ \cline{2-7}

&   & 0 & 26  & 78& 25 & 74 \\ 
& 4$\rho_0$ & 0.3 & 37 & 111 &  40 &121 \\  
& &0.5&45&136&48&148\\ \hline
\end{tabular}
\caption{Table shows the effect of baryonic density $\rho_B$ and isospin asymmetric parameter $\eta$ on the shift in decay constants (in MeV) of $D_{1S}$ and $B_{1S}$ mesons for different values of strangeness fraction $f_s$}
\label{D1sdecayshifttable}
\end{table}

The mass shift and shift in decay
constants of charmed and bottom vector and axial-vector
mesons have been investigated in past using QCD sum rules
in symmetric nuclear matter
  only \cite{wang2, wang3}  . The values of mass shift
  for vector mesons $D^{*}$ and $B^{*}$
  in leading order (next to leading order) calculations were 
    -70 (-102) and -340 (-687) MeV respectively.
    For the axial vector $D_{1}$ and $B_{1}$ mesons
    the above values of mass shift changes
    to 66 (97) and 260 (522) MeV respectively.
    The values of shift in decay constant for
    $D^{*}$ and $B^{*}$ mesons in
    leading order (next to leading order) 
     are found to be -18(-26) and -55(-111) MeV, whereas,
    for $D_{1}$ and $B_{1}$ mesons 
    these values changes to 21(31)
     and  67 (134) MeV respectively.
    We can compare the above values of
    mass shift (decay shift) to our results -63 (-20), -312 (-48.9), 62 (20)
     and 216 (53) MeV 
    for $D^{*}$, $B^{*}$, $D_{1}$ and $B_{1}$ mesons
    evaluated using $m_u$ = $m_d$ = 7 MeV 
    in symmetric nuclear matter ($\eta$ = 0 and $f_s$ = 0).
    The observed negative values of mass shift for vector $D^{*}$  and $B^{*}$
mesons
in nuclear and strange hadronic matter favor the
decay of higher charmonium and bottomonium states to $D^{*}\bar{D}^{*}$ and
$B^{*}\bar{B}^{*}$
pairs and hence may cause the quarkonium suppression.
However the axial-vector meson undergo a positive mass
shift in nuclear and strange hadronic medium and 
hence the possibility  of  decay of excited charmonium and bottomnium
states to $D_{1}\bar{D}_{1}$ and $B_{1}\bar{B}_{1}$
pairs is suppressed. 
The observed effects of isospin asymmetry of the
medium on the mass modifications of
$D$ and $B$ mesons can be verified
experimentally through the ratios
$\frac{D^{*+}}{D^{*0}}$, $\frac{B^{*+}}{B^{*0}}$,
$\frac{D_{1}^{+}}{D_{1}^{0}}$ and 
$\frac{B_{1}^{+}}{B_{1}^{0}}$ whereas the
effects of strangeness of the matter
can be seen through the 
ratios $\frac{D^{*}}{D_{S}^{*}}$, $\frac{B^{*}}{B_{S}^{*}}$,
$\frac{D_{1}}{D_{1S}}$ and $\frac{B_{1}}{B_{1S}}$.
The traces of observed medium modifications
of masses and decay constants
can be seen experimentally
in the strong decay  width and leptonic decay width of
heavy mesons 
\cite{azzi}. For example, in ref. \cite{bel1}
the couplings $g_{D^{\star}D \pi}$ and
$g_{B^{\star}B \pi}$ were studied
using the QCD sum rules and 
strong decay width of charged vector $D^{\star +}$ mesons
for the strong decay, $D^{*+}$ $\longrightarrow$ $D^{0} \pi^{+}$
were evaluated using the formula,
\begin{equation}
\Gamma(D^*\longrightarrow D\pi)=\frac{g^2 _{D^*D \pi}}{24 \pi m^2 _{D^*}} |k_{\pi}|^3,
\label{decayw1}
\end{equation}
where pion momentum, $k_{\pi}$ is, 
\begin{equation}
 k_{\pi} = \sqrt{\frac{(m_{D^*} ^2 - m_{D}^2 + m_{\pi}^2)^2}{(2m_{D^*})^2} - m_\pi ^2}.
 \label{decayw2}
\end{equation}
In equation (\ref{decayw1}) coupling $g^2 _{D^*D \pi} = 12.5\pm1$,
$m_{D^{\star}}$ and $m_D$ denote the masses of
vector and pseudoscalar meson respectively.
From equation (\ref{decayw1}) we observe that
the values of decay width depend upon the  masses of 
vector mesons, $D^{\star}$ and pseudoscalar $D$ 
mesons.
Using vacuum values for the masses of $D$ mesons the values of decay width, 
$\Gamma(D^{*+}$ $\longrightarrow$ $D^{0} \pi^{+})$ 
are observed to be $32\pm5$ keV \cite{bel1}. 
However, as we discussed in our present work the
 charmed mesons
get modified in the hadronic medium and this must
lead to the medium modification of decay width
of these mesons.
For example, if we 
consider the in-medium masses of vector
mesons from our present work and for
pseudoscalar mesons
we use the in-medium masses from 
ref. \cite{amavstranged} (in this reference the
mass modifications of pseudoscalar $D$ mesons were
calculated using the chiral SU(4) model in nuclear
and strange hadronic medium), then in nuclear medium $(f_s = 0)$,
at baryon density, $\rho_B$ = $\rho_0$,
the values of decay width,
$\Gamma(D^{*+}$ $\longrightarrow$ $D^{0} \pi^{+})$, 
are observed to be $219$ keV and $31$ keV
at asymmetry parameter $\eta = 0$ and $0.5$ respectively.
In strange medium($f_s = 0.5$), the
above values of decay width will change to
$84$ and $25$ keV at $\eta = 0$ and $0.5$
respectively. We observe that the decay width of heavy mesons
vary appreciably because of
medium modification of heavy meson masses. In our future work we shall
evaluate in detail the effects of medium modifications
of masses and decay constants of
heavy charmed and bottom mesons on the
above mentioned experimental observables.
Also the effects of finite temperature 
of the strange hadronic medium
on the properties of vector and axial-vector mesons 
will be evaluated.

\section{Summary}
In short, we computed the 
 mass shift and shift of decay constants of vector
and axial vector charm and bottom mesons in 
asymmetric hadronic matter,  consisting of nucleon and
hyperons, using
phenomenological chiral model and QCD sum rules.
For this, first the 
quark and gluon condensates were calculated using
chiral hadronic model and then using these values
of condensates as input in QCD sum rules
 in-medium properties 
of vector and axial-vector mesons
were evaluated. We observed a negative   (positive)  shift  in the masses and decay 
constants of
  vector (axial vector) mesons.
  The magnitude of shift increases with increase in 
  the density of baryonic density of matter.
The properties of mesons are
seen to be sensitive for the isospin asymmetry 
as well as strangeness fraction of the 
medium.  The isospin asymmetry of the
medium causes the mass-splitting 
between isospin doublets, whereas, the
the presence of hyperons
in addition to nucleons
lead to an increase
in the magnitude of shift in the masses and decay constants
 of  heavy mesons.  
The observed effects on the masses and decay 
constants of heavy vector and axial vector
mesons may be reflected  
 experimentally
in  the production ratio of open charm mesons as well as in their decay width.
The negative mass shift of
charmed vector mesons as observed
in present calculations may cause the
formation of bound states
with the nuclei
as well as the decay of excited charmonium 
states to $D^{*}\bar{D}^{*}$ pairs causing
charmonium suppression.
The present work on the in-medium mesons properties
may be helpful in understanding the
experimental observables
of CBM and PANDA experiments of
FAIR project at GSI Germany.

\acknowledgements 
 The authors gratefully acknowledge the financial support from
the Department of Science and Technology (DST), Government of India for research project under
 fast track scheme for young scientists (SR/FTP/PS-209/2012).


\begin{thebibliography}{1}
\bibitem{cleo} http://www.lepp.cornell.edu/Research/EPP/CLEO/
\bibitem{bele} http://belle.kek.jp/
\bibitem{babar} http://www-public.slac.stanford.edu/babar/
\bibitem{qmc1}
  K. Tsushima, D.H. Lu, A.W. Thomas, K.Saito
and R.H. Landau, Phys. Rev. C {\bf 59}, 2824 (1999).
\bibitem{tolos2}
  L. Tol\'os, D. Cabrera and A.  Ramos, Phys. Rev. C {\bf 78}, 045205 (2008).
  \bibitem{tolos3}
  L. Tolos, C. Garcia - Recio, and J. Nieves, Phys. Rev. C {\bf 80}, 065202 (2009).
  \bibitem{tolos4}
Laura Tol\'os, Raquel Molina, Daniel Gamermann and Eulogio Oset,
  Nucl. Phys. A {\bf 827}, 249c (2009).  
  \bibitem{higler1}
 T. Hilger,R. Thomas, B. Kampfer, Phys. Rev. C {\bf 79}, 025202 (2009).
 \bibitem{haya1}
 Arata Hayashigaki, Phys. Lett. B {\bf 487}, 96 (2000).
 \bibitem{wang1}
 	Zhi-Gang Wang, Tao Huang, Phys. Rev. C {\bf 84},  048201  (2011).
\bibitem{wang2}
 	Zhi-Gang Wang, Int. J. Mod. Phys. A {\bf 28}, 1350049 (2013) .
\bibitem{paper3}
 	P. Papazoglou, D. Zschiesche, S. Schramm, J. Schaffner - Bielich, H. Stocker and W. Greiner, Phys. Rev. C {\bf 59}, 411 (1999).
 \bibitem{amarind} 
Amruta Mishra and Arindam Mazumdar, 
Phys. Rev. C {\bf 79},  024908 (2009).	
 \bibitem {amdmeson} 
A. Mishra, E. L. Bratkovskaya, J. Schaffner-Bielich, 
S.Schramm and H. St\"ocker, Phys. Rev. C {\bf 69}, 015202 (2004). 	
\bibitem{amarvind} 
Arvind Kumar and Amruta Mishra, Phys, Rev. C {\bf 81}, 065204 (2010).
\bibitem{amavstranged} 
Arvind Kumar and Amruta Mishra , Eur. Phys, J. A {\bf 47}, 164 (2011). 	
 	
 	\bibitem{tolosbnd}
   L. Tol\'os, C. Garcia-Recio, J. Nieves, O. Romanets, L. L. Salcedo, Few-Body Syst {\bf 54}, 923 (2013).
   \bibitem{tolos5}
   C. Garcia  Recio, J. Nieves, L. Tolos, Phys. Lett. B  {\bf 690}, 369 (2010).

 	
 \bibitem{arv1}
 	Arvind Kumar, Adv. in High Energy Physics {\bf 2014},  549726 (2014) .	
 	
\bibitem{ebert1}
D. Ebert, R.N. Faustov and V.O Galkin, 
Mod. Phys. Lett. A {\bf 17}, 803 (2002). 	
 	
 \bibitem{bel1}
V. M. Belyaev, V. M. Braun, A. Khodjamirian and R. Ruckl, Phys. Rev. D {\bf 51},
6177 (1995).	
\bibitem{shurya1}
E. V. Shuryak, Nucl. Phys. B {\bf 198}, 83 (1982).
\bibitem{heq1}
M. Neubert, Phys. Rev. D {\bf 45},  2451 (1992).
\bibitem{heq2}
E. Bagan, P. Ball, V. M. Braun and H. G. Dosch, Phys. Lett. B {\bf 278}, 457 (1992).
\bibitem{gel2}
D. J. Broadhurst, Phys. Lett. B {\bf 101}, 423 (1981).
\bibitem{gel3}
S. C. Generalis, J. Phys. G {\bf 16}, 785 (1990).
\bibitem{gel4}
C. A. Dominguez and N. Paver, Phys. Lett. B {\bf 246}, 493 (1990).
\bibitem{gel1}
P. Gelhausen, A. Khodjamirian, A.A. Pivovarov and D. Rosenthal, Phys. Rev. D 
{\bf 88}, 014015, (2013).
\bibitem{wang3}
Zhi-Gang Wang, arXiv:1501.05093 [hep-ph].
\bibitem{azzi}
K. Azizi, N. Er,
H. Sundu, Eur. Phys. J. C {\bf 74}, 3021 (2014).
\bibitem{azzi2}
E Veli Veliev et.al., J. Phys.: Conf. Ser.  {\bf 347}, 012034 (2012).
\bibitem{isoamss} 	
A. Mishra and S. Schramm, Phys. Rev. C {\bf 74}, 064904 (2006),	
A. Mishra, S. Schramm and W. Greiner, Phys. Rev. C {\bf 78}, 024901 (2008) and H. St\"ocker, and W. Greiner, Phys. Rev. C {\bf 59},  411  (1999).	
\bibitem{heide1}
Erik K. Heide, Serge Rudaz and Paul J. Ellis, Nucl. Phys. A {\bf 571}, 
 713 (2001).
 \bibitem{cohen}
Thomas D. Cohen, R. J. Furnstahl and David K. Griegel, Phys. Rev. Lett. {\bf 67}, 961 (1991).
\bibitem{koike1}
Yuji Koike and Arata Hayashigaki, 
 Prog. Theo. Phys. {\bf 98}, 631 (1997).
 
 \bibitem{kwon1}
   Youngshin Kwon, Chihiro Sasaki, Wolfram Weise,
    Phys. Rev. C {\bf 81}, 065203 (2010).
    
     \bibitem{zscho1}
   S. Zschocke, O.P. Pavlenko, B. Kampfer,
   Eur. Phys. J. A {\bf 15}, 529 (2002).  
 
 
 
\bibitem{qcdThomas}
R. Thomas, T. Hilger, B. Kampfer, Nucl. Phys. A {\bf 795}, 19 (2007).


 	
 \bibitem{quark2}
M. Lutz, S. Klimt and W. Weise,
 Nucl. Phys. A {\bf 452}, 521 (1992).
 \bibitem{quark3}
G.Q. Li, C.M. Ko, Phys. Lett. B {\bf 338}, 118 (1994).
\bibitem{quark4}
Delfino, A. , Dey, J. , Dey, M. et al., Phys. Lett. B {\bf 363}, 17 (1995).
\bibitem{quark5}
Guo, H., J. Physics (London) G. {\bf 25}, 1701 (1999).
\bibitem{quark6}
K. Saito, K. Tsushima, A. W. Thomas, Mod. Phys. Lett. A {\bf 13}, 769 (1998).

  
	 
\end{thebibliography}
\end{document}